\documentclass{article}
\usepackage{multirow}
\usepackage{microtype}
\usepackage{graphicx}
\usepackage{subfigure}
\usepackage{booktabs} 
\usepackage{xcolor}
\usepackage{threeparttable}
\usepackage{amsmath}
\usepackage{soul}
\usepackage{amsfonts}
\usepackage{hyperref}
\usepackage{xurl}
\usepackage{dblfloatfix}
\usepackage{multirow}
\usepackage{tabularx}
\usepackage{algorithm}
\usepackage{tikz}
\newcommand*\circled[1]{\tikz[baseline=(char.base)]{
            \node[shape=circle,draw,inner sep=0.75pt] (char) {\footnotesize #1};}}

\usepackage[accepted]{template}

\templatetitlerunning{HELIX: HEterogeneous LLM Inference with FleXible and Fine-Grained Partitioning}
\graphicspath{{template/}}
\begin{document}

\twocolumn[
\templatetitle{HELIX: HEterogeneous LLM Inference with FleXible and Fine-Grained Partitioning}



\templatesetsymbol{equal}{*}

\begin{templateauthorlist}
\templateauthor{Victoria Clerico}{ibm,tud}
\templateauthor{William Simon}{ibm}
\templateauthor{Corey Liam Lammie}{ibm}
\templateauthor{Orhun Görkem}{ibm}
\templateauthor{Athanasios Vasilopoulos}{ibm}
\templateauthor{Hsinyu Tsai}{ibm-us}
\templateauthor{Jeronimo Castrillon}{tud}
\templateauthor{Abu Sebastian}{ibm}
\templateauthor{Hadjer Benmeziane}{ibm}
\end{templateauthorlist}
\templateaffiliation{ibm}{IBM Research–Zurich Research Lab, Rüschlikon, Switzerland}
\templateaffiliation{ibm-us}{IBM Research–Silicon Valley Lab, San Jose, CA, USA}
\templateaffiliation{tud}{Technische Universität Dresden, Dresden, Germany}
\templatecorrespondingauthor{Victoria Clerico}{mvc@zurich.ibm.com}

\templatekeywords{LLM inference, heterogeneous platforms, disaggregation, model partitioning}
\vspace{5pt}
\begin{abstract}
LLM inference is inherently heterogeneous, with performance bottlenecks shifting throughout inference phases down to individual operators. 
This has motivated heterogeneous serving systems that disaggregate execution across specialized accelerators, typically through fixed partitions such as prefill/decode or attention/FFN. 
However, these approaches commit to a partitioning strategy in advance, limiting their ability to adapt to different workloads and model architectures while forcing emerging accelerators into predefined roles.
We present HELIX, a vLLM-based software stack that discovers, compiles and dispatches workload-aware partitions across heterogeneous accelerators at the operator granularity.
At engine initialization, HELIX automatically determines hardware-specific operator placements that minimize execution latency, accounting for profiled execution costs, communication overheads, device capabilities, and memory constraints. 
It combines dynamic programming, piecewise compilation, and heterogeneous subgraph dispatch, enabling accelerators with partial operator coverage to participate in LLM inference without predefined model partitions.
HELIX provides an open-source unified framework for exploring, evaluating, and integrating heterogeneous accelerators for LLM inference, lowering the barrier to emerging hardware by exposing them as subgraph executors.
In systems combining H100 and B200 GPUs with emerging accelerators, HELIX resolves LLM partitions that multi-query decoding latency by up to 3$\times$ for dense and sparse transformer models and up to 3.5$\times$ for hybrid state-space models, relative to standard attention-FFN disaggregation. 
\end{abstract}
]




\printAffiliationsAndNotice{}  

\section{Introduction}

The rapid adoption of LLMs for conversational assistants, code synthesis and complex reasoning tasks has made inference efficiency a first-order systems concern. 

\begin{figure*}[t]
    \centering
\includegraphics[width=0.9\textwidth]{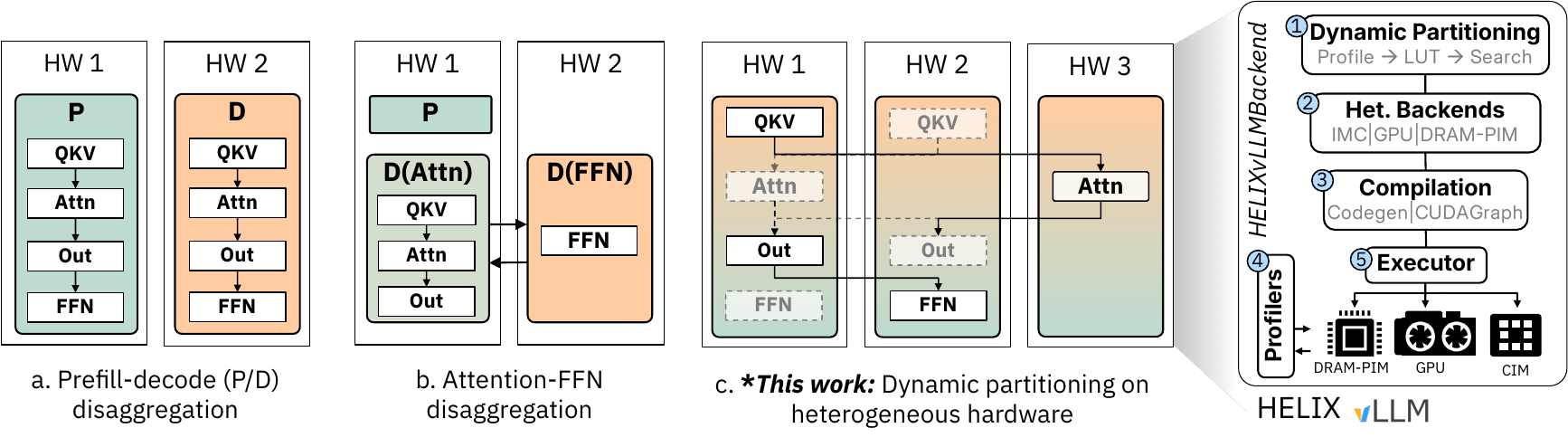}
\caption{LLM inference disaggregation. (a) P/D assigns each phase to separate hardware. (b) A/F further splits decode attention and FFN. (c) HELIX maps operators across heterogeneous hardware (HW 1--3) at engine initialization via \circled{1} profiling-based partition search, \circled{2} heterogeneous backends, \circled{3} piecewise compilation, and \circled{4}-\circled{5} heterogeneous dispatch with per-device profiling.}
\label{fig:Figure1}
    \vspace{-12pt}
\end{figure*}

Achieving efficient hardware mapping requires recognizing that LLM inference is far from monolithic, as the hardware demands it imposes vary substantially depending on which stage of the inference pipeline is examined. 
At the phase level, \textit{prefill} processes the input prompt in parallel and is predominantly compute-bound, whereas \textit{decode} generates tokens in an auto-regressive fashion and is typically bandwidth-bound.
This asymmetry has motivated prefill/decode (P/D) disaggregation~\cite{qin2025mooncakekvcachecentricdisaggregatedarchitecture, Zhong2024DistServe}, which assigns each phase to a dedicated hardware pool optimized for its respective bottleneck (see Figure~\ref{fig:Figure1}a).
Within the decode phase itself, a similar asymmetry emerges at the operator level: for high-concurrency multi-query serving, feed-forward networks (FFN) amortize the large set of weight parameters fetched from memory to generate multiple tokens. Instead, attention is strictly memory-bound as each new token must attend over the entire growing KV-cache, requiring repeated memory reads that scale with sequence length rather than reuse of loaded data.
This operator-level asymmetry has encouraged Attention--FFN (A/F) disaggregated systems~\cite{step3system, liang2025injectingadrenalinellmserving, Zhu2025MegaScale,nvidiaInsideNVIDIA}, which map attention and FFN or expert layers onto separate resources (see Figure~\ref{fig:Figure1}b).
However, even these operator-level bottlenecks are not invariant.
At the workload level, batch size modulates the arithmetic intensity of FFN kernels during decode~\cite{PAPI2025asplos}, since larger batches amortize the loaded weights. 
Therefore, variations in the serving scenario can cause individual kernels to transition between memory- and compute-bound regimes, making them better suited to different hardware architectures.
This heterogeneity has led to the development of disaggregated inference systems, with A/F emerging as a prominent direction. 
Beyond GPU-only deployments, the introduction of specialized accelerators has also encouraged heterogeneous disaggregation, since hardware properties favor specific operators, such as LPU-based FFN offloading from GPU~\cite{nvidiaInsideNVIDIA} and DRAM process-in-memory (PIM) for memory-bound attention~\cite{Park2024AttAcc, PAPI2025asplos, asplos2026REPA}.
These heterogeneous systems commonly use model-specific partition boundaries and focus the disaggregation decision on the sharding exploration rather than jointly deciding fine-grained placements down to the individual operators. 
The disaggregation strategy is inherited from the A/F framing and never revisited against the particular model, hardware combination and serving scenario.
As hardware architectures diversify, fixed partition boundaries cannot account for the trade-offs introduced by combining different hardware capabilities, adopting new model structures, and varying serving regimes.
This paper argues that LLM serving must be rebuilt around \emph{dynamic, fine-grained partitioning}.
Here, dynamic denotes deployment-specific placement resolved during compilation time.
\textit{What} to offload and \textit{where} to place it are decision variables searched over the LLM computational graph and resolved against the hardware's capabilities, interconnect and computation cost for a model and a target serving regime.
We formulate partitioning as a dynamic programming problem over the LLM computational graph.
Using profiled per-operator execution and communication costs, our method finds latency-minimizing partitions for a given model, hardware configuration, and serving regime.
We present \textbf{HELIX}, a heterogeneous execution stack built on vLLM~\cite{kwon2023efficient}, which compiles and executes partitioned LLM graphs across available accelerators.
Figure~\ref{fig:Figure1} places our approach in this landscape, moving from static P/D and A/F toward the dynamic partitioning that accommodates workloads, models, and hardware pool.
In summary, this paper makes the following contributions:
\setlength{\itemsep}{0pt}
\setlength{\parskip}{0pt}
\setlength{\topsep}{0pt}
\begin{itemize}
    \item We develop a profile-guided \textbf{dynamic programming} method for partitioning search to determine, at compilation time, how to split LLM inference stages across heterogeneous hardware, given workload, system, and model.
    \item We design HELIX, a heterogeneous execution stack built on vLLM, that along with resolving partitions, runs LLM inference across the available devices, backed by pluggable per-device profilers and backends.
    \item We evaluate HELIX on dense Transformer, MoE, and hybrid SSM models across four hardware platforms and four heterogeneous configurations where found partitions can reduce multi-query decoding latency up to 3.5$\times$ compared to attention-FFN disaggregation.
\end{itemize}
\section{Background}\label{sec:background}
\paragraph{\textbf{Disaggregated serving}:}
P/D disaggregation assigns prefill and decode to separate compute resources connected by KV-cache transfers~\cite{Zhong2024DistServe} (see Figure~\ref{fig:Figure1}a).
Heterogeneous GPU cluster variants extend this idea by routing prefill to high-FLOPS devices and decode to high-bandwidth ones~\cite{Patel2024SplitwiseISCA, Jiang2025hexgenICLR}.
A/F further separates attention and FFN or expert layers into dedicated GPU pools~\cite{lin2024infinitellmefficientllmservice, step3system, Zhu2025MegaScale} (Figure~\ref{fig:Figure1}b).
At the orchestration layer, NVIDIA Dynamo~\cite{Pei2026Dynamo} realizes these ideas by providing request routing, KV-cache-aware scheduling, and cross-engine coordination for disaggregated deployments. 
Frameworks such as AIConfigurator~\cite{aiconfigurator} optimize resource allocation and sharding settings for a given workload \footnote{AIConfigurator added A/F support in a later commit (\url{https://github.com/ai-dynamo/aiconfigurator/pull/1597})}.
However, this remains an offline \textit{sizing} problem over GPU resources rather than a search for operator placement across heterogeneous accelerator types.
\paragraph{\textbf{Heterogeneous hardware for inference:}}
Beyond GPU-only disaggregation, recent systems pair compute-centric and memory-centric devices within a single inference pipeline for A/F.
Attention-centric GPU-PIM designs place attention on PIM and retain other operators on the GPU~\cite{Park2024AttAcc, Heo2024NeuPIMASPLOS,asplos2026REPA}. 
PIM offloading also extends to state updates in SSMs~\cite{PIMBA}. 
More recently, NVIDIA’s proposed a complementary design that keeps attention on GPUs (Vera Rubin) and offloads static FFN and MoE expert layers to Groq LPUs~\cite{nvidiaInsideNVIDIA}, reporting substantial throughput gains over GPU-only execution.
Similarly, emerging 3D compute-in-memory (3DCIM) hardware can heavily accelerate memory-bound matrix multiplications, and has been paired with DPUs for attention to avoid costly weight transfers~\cite{ICCD20253dnand, ISCA263dnandMoE, Bchel2025Efficient}.
\section{Motivation}\label{sec:motivation}
\begin{figure}[ht]
    \centering
    \vspace{-10pt}
\includegraphics[width=0.475\textwidth]{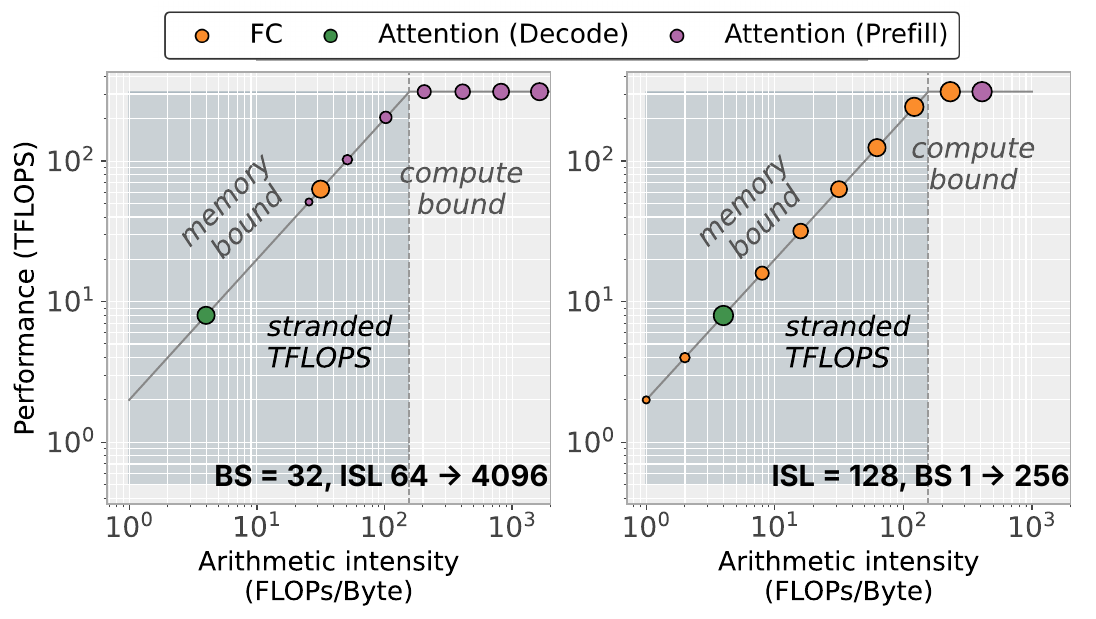}
\vspace{-15pt}
        \caption{Roofline analysis on an A100-80GB showing workload-dependent operator bottlenecks.
Marker size increases with input sequence length (ISL, \textbf{Left}) and batch size (BS, \textbf{Right}).}
        \vspace{-15pt}
    \label{fig:Figure2}
\end{figure}

\paragraph{Challenges in Attention--FFN disaggregation:}
Figure~\ref{fig:Figure2} illustrates workload-dependent shifts in operator bottlenecks on an A100-80GB: increasing batch size moves decode general matrix multiplications (GEMMs) from memory- to compute-bound.
Recent analytical work on A/F provisioning confirms this at the system level, showing that the optimal attention-to-FFN device ratio depends on stochastically evolving KV-cache occupancy and number of requests~\cite{song2026analytical}.
Whether an operator is compute- or memory-bound does not alone determine where it should run. 
Optimal operator placement also depends on the hardware combination, and on which regimes each device performs best, since in heterogeneous systems, platforms are paired to complement each other and usually differ in the operators and regimes they execute efficiently.
Other factors, including interconnect bandwidth, memory capacity, energy, and cost per token, further shape operator placement, which a fixed attention-FFN split cannot capture.
\paragraph{Limitations of serving engines on heterogeneous platforms.}
Existing A/F solutions~\cite{step3system, nvidiaInsideNVIDIA, Zhu2025MegaScale, liang2025injectingadrenalinellmserving} converge on a common design, where an orchestrator assigns hardware pools and sharding configurations to prefill, attention, and FFN, each served by an independent, network-connected engine.
Supporting a new accelerator therefore means porting the entire engine to it, which renders it impractical for emerging platforms and infeasible for specialized attention or GEMM units.

HELIX instead intervenes one layer below, at the \emph{engine} level, treating heterogeneous placement as a \emph{data-plane} rather than an orchestration problem.
Each accelerator acts as a subgraph executor rather than hosting an independent serving engine, using a \texttt{torch.compile}-compatible backend to execute its supported operations.
This design suits tightly interconnected single-node heterogeneous systems that lack the device pools and datacenter fabric that justify one engine per operator class as in existing methods \cite{liang2025injectingadrenalinellmserving,aiconfigurator,step3system}.

\section{HELIX: Profile-guided Partitioning Search}\label{ref:dp}
HELIX comprises two components that interface through the partitioning choice but that can be used independently.
The first is a profile-guided partition search that, given a model, serving regime and hardware combination, maps the model's main computational primitives (GEMMs, attention, state scan and update, etc.) to available hardware, minimizing end-to-end latency under memory capacity constraints.
The second is a vLLM-based execution stack that runs the model under a given placement (Section~\ref{sec:vllm_HELIX}).
The search can therefore be replaced by a heuristic-based partition, and the stack can serve on its own as a heterogeneous model executor within a practical inference engine.

The partition search operates on the model's \texttt{torch.fx} graph, which contains on the order of a thousand nodes for models like Llama-8B (Figure \ref{fig:graph-example}(a-b)).
Only a few of these nodes carry the non-negligible compute cost of the LLM computational graph (i.e., the computational primitives).
We declare these \textit{placeable} on a set of candidate devices defined by a compatibility map (e.g., \texttt{QKV} $\rightarrow$ \texttt{HW1}, \texttt{HW2}).
The search omits the costs of the remaining element-wise and metadata operations (reshape, view, transpose, slice, etc.) as they typically account for a small percentage of the execution time (see Appendix Figure \ref{app_fig:exec_time_per_model}), while full graph execution and evaluation include these element-wise costs.
These operations are absorbed into their producers only when supported by the assigned backend.
Otherwise, they remain explicit placeable stages with their transfer dependencies but bear no cost (e.g., \texttt{SiLU} $\rightarrow$ \texttt{HW1}).
Figure \ref{fig:graph-example}c illustrates the reduced directed acyclic graph (DAG).

The reduced DAG has $N$ stages, proportional to the number of layers and the number of \textit{placeable} primitives per layer.
Each primitive is profiled once, offline, over the shapes it takes under the evaluated workloads, so scoring a candidate placement costs a table lookup in microseconds rather than a deployment in minutes.
Recent configuration-search tools adopt the same abstraction~\cite{aiconfigurator}, which holds as long as a primitive's isolated latency predicts its runtime latency.
This is the case on GPUs, where consecutive kernels execute with little cross-kernel pipelining (see Appendix Figure~\ref{app:fusion_gpu}).
Our implementation can also use roofline-based cost estimates instead of profiled costs, but we omit this from our evaluation since roofline models are generally less accurate than profiled data (see Appendix Figure \ref{app_fig:roofline_vs_profile}).

Placing a computation graph onto heterogeneous devices is NP-complete in general~\cite{HeterogeneousDAGScheduling}, so practical methods exploit model structure. Following prior device-placement work~\cite{PipeEdgeDP}, HELIX adopts a dynamic programming (DP) formulation and extends it to the residual dependencies and memory constraints in LLM stage graphs, keeping the search within the time budget of compilation time during the vLLM engine initialization.
\subsection{Placement and objective}
\begin{figure}[!ht]
    \centering
\includegraphics[width=0.475\textwidth]{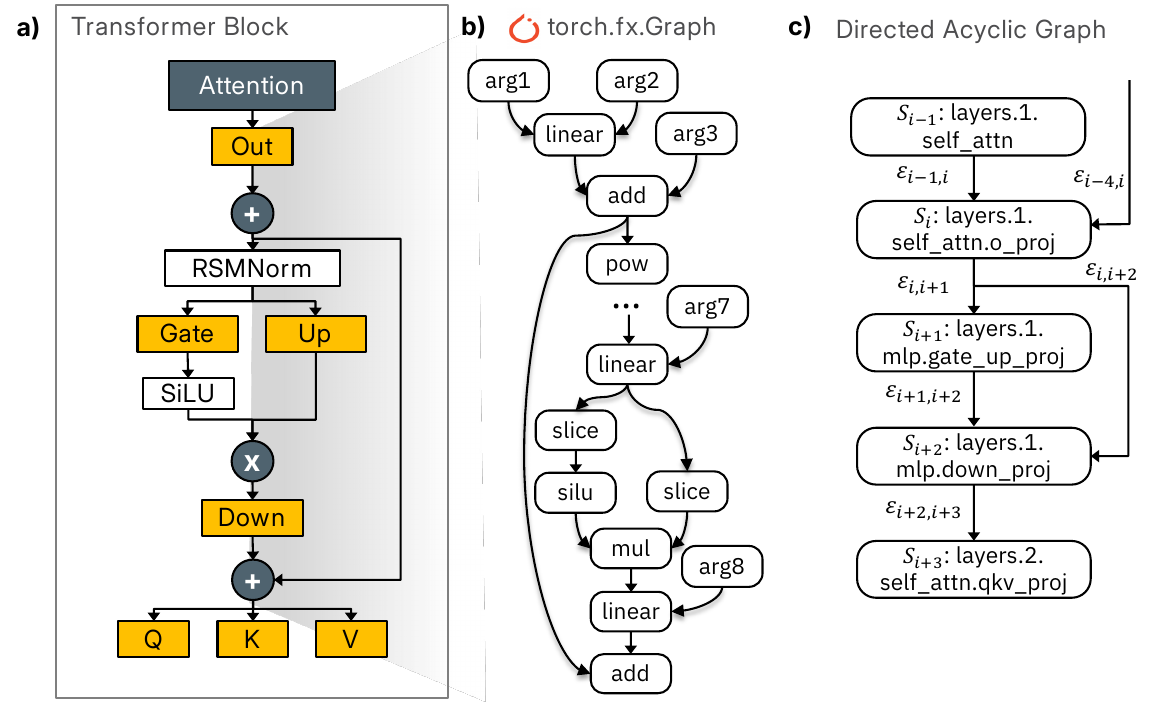}
\vspace{-10pt}
    \caption{Graph reduction from (a) a transformer block through (b) its torch.FX graph to (c) a placement DAG, preserving sequential backbone and residual dependencies.}
    \label{fig:graph-example}
\end{figure}

We represent the reduced computation graph as a DAG $G=(\mathcal{S},\mathcal{E})$, with stages $\mathcal{S}=\{s_0,\dots,s_{n-1}\}$ in topological order.
In standard chain-like DAGs, every stage $s_i$  $(i>0)$, consumes the output of $s_{i-1}$. 
However, residual additions in LLMs create nonadjacent dependencies as shown in Figure~\ref{fig:graph-example}, so a stage may have multiple predecessors, denoted $\mathrm{pred}(i)$. 
Each stage $(s_i)$ has an admissible device set $(C_i)$ and each device $(d)$ has available memory ($M_d$).
Assigning $(s_i)$ to $d$ has a computational cost $(T_{\mathrm{comp}}(s_i, d))$ and a memory footprint $(m(s_i))$, comprising parameters, KV cache, or recurrent state.
The budget $M_d$ is computed after reserving runtime headroom.
A graph edge $(s_j,s_i)\in\mathcal{E}$ carries $w_{ij}$ elements of $B$ bytes.
When $s_j$ and $s_i$ are placed on $d'$ and $d$, the transfer cost $T_{\mathrm{comm}}(w,d,d') = w \cdot B\,/\,\beta_{d,d'} + \sigma$ where $B$ is the number of bytes per element and $\beta_{d,d'}$ is effective directed-link bandwidth, plus $\sigma$ a fixed startup latency, excluding buffer-capacity effects.

Based on these definitions, a \emph{placement} $a:\{0,\dots,n-1\}\rightarrow\mathcal{D},\qquad a(i)\in C_i$ assigns each stage to an admissible device (i.e., a partition of the DAG up to stage $s_{i}$).
It determines where each operator (or set of operators) executes and which intermediate tensors cross a hardware link and, hence, the latency of the forward pass.

We define the finish time of stage $s_i$ as the arrival time $A_i(a)$ of its latest-arriving input plus the execution time of the cost-bearing operation of the stage,
\begin{equation}
\label{eq:latency}
L_i(a)=A_i(a)+T_{\mathrm{comp}}\big(s_i,a(i)\big).
\end{equation}
How arrivals from multiple inputs combine depends on where their producers execute.
Host inputs are treated as produced on $d_{\mathrm{init}}$.
For $d=a(i)$, the output of a predecessor $s_j$, $j\in\mathrm{pred}(i)$, is available at $L_j(a)$ if also $s_j$ runs on $d$, and inputs already transferred to $d$ for an earlier consumer are reused without a new transfer.
The remaining inputs, produced on a remote device $d'\neq d$, contend for the link $(d',d)$, whose link state $Q_{d' \to d}$ (initially $0$) must record when it frees.
Each link serves transfers in the order their data is consumed, and transfers feeding the same stage, in the order they become available.
The $k$-th transfer completes at:
\begin{equation}
\label{eq:arrival-link}
A_{i,d'}(k)=\max\big(L_{j_k}(a),\,A_{i,d'}(k-1)\big)+T_{\mathrm{comm}}\big(w_{j_k i},d,d'\big),
\end{equation}
Let $A_{i,d'}$ denote the time at which the transfer of the last input of $s_i$ sent over link $(d',d)$ finishes.
The link then remains occupied until this time, so we set $Q_{d'\to d}\leftarrow A_{i,d'}$ for the next stage to account for link contention.
The first term of the $\max$ in Eq. \ref{eq:arrival-link} assumes a transfer starts once its input is released, so it cannot overlap with the producer's own subgraph execution. 
However, execution of stages on the destination device $d$ can overlap with transfers $d'\to d$.
For example, a residual input to $s_i$ can be transferred while $s_{i-1}$ computes on the same $d$.
The second term makes the transfer wait for earlier transfers on the same link.
Links between distinct device pairs proceed concurrently, so the arrival time of the last input is:
\begin{equation}
\label{eq:arrival}
A_i(a)=\max\Big(\max_{j\in\mathrm{pred}(i):\,a(j)=d}L_j(a),\ \max_{d'\neq d}A_{i,d'}\Big)
\end{equation}
Since arrivals combine by $\max$, a transfer adds latency only if it delays the last input of some stage, either directly or by occupying the link for later transfers.

Considering the previous equations, the best candidate placement (i.e. partition) is the one that minimizes the finish time of the last stage, 
\begin{equation}\label{eq:objective}
    a^{*}=\arg\min_{a}L_{n-1}(a)
\end{equation}
and is memory-feasible when the footprint it accumulates on each accelerator stays below $\sum_{i\,:\,a(i)=d} m(s_i) \;\le\; M_d$
\subsection{Partitioning search via dynamic programming}\label{sub:dp}
The objective in Eq. \ref{eq:objective} defines a space of possible placements of size $D^{N}$ for a model with $N$ stages and up to $D$ admissible devices, and every candidate must additionally be checked against the per-device memory budgets.
Collecting profiling data makes each placement candidate cheap to score, a table lookup rather than a deployment, but does not make exhaustive search tractable.
HELIX instead uses dynamic programming along the sequential backbone, retaining the earlier outputs needed by residual dependencies and the memory committed on capacity-constrained devices.
We extend stage-indexed DP formulations \cite{PipeEdgeDP} with these two state components, described below; Appendix C provides the full algorithm.
\subsubsection{The frontier state}
In LLM DAGs, fan-in and fan-out stages arise from the residual connections within and between layers (see Figure~\ref{fig:graph-example}).
Therefore, the optimal placement of a fan-in stage $i$ does not depend only on the immediate predecessor stage $i-1$ (as DP solution could track) but must be resolved based on all possible arrivals from all of its producers.
We therefore extend the standard DP state to retain all stages within a fan-out$\to$fan-in structure.
For instance, if $s_{i}$ is a fan-in and consumes both $s_{i-1}$ and $s_{i-2}$, and, $s_{i-1}$ also consumes $s_{i-2}$, a partial placement that minimizes the completion time of $s_{i-1}$ may not minimize that of $s_i$, so states of $s_{i-2}$ and $s_{i-1}$ shall not be resolved until $s_{i}$ is evaluated.
Under Equations \ref{eq:arrival-link}--\ref{eq:arrival}, a partial placement now depends on predecessor assignments and finish times together with the directed-link availability $Q$.
Outside the fan-out and fan-in, every stage consumes at most the output of its predecessor, where a single transfer is in flight at a time per link. For a fan-in stage $s_i$, link contention can arise between the residual skip and a transfer feeding an intermediate stage $s_{i-1}$. 
Link becomes available by the time $s_{i-1}$ starts executing, which lets us simplify $Q_{d'\to d}=A_{i-1}$ when $s_{i-1}$ runs on $d$, and $0$ otherwise.
This simplification relies on our LLM DAG abstraction and general DAGs, since they can have many transfers in flight at once, would require tracking $Q$ explicitly in the state.
A residual input to $s_i$ is then sent after the inputs of $s_{i-1}$ (since they are consumed earlier) and its transfer overlaps with $s_{i-1}$ computation.
We call the set of stages between a fan-out and its fan-in the \textit{frontier}: it opens when the residual producer is placed and closes once the fan-in consumes its output.
The DP is no longer indexed by stage and device $(i,d)$ but by the joint placement of the stages within the frontier.

Under each frontier assignment, the DP state retains one or more labels $\ell$, i.e., partial placements that share the frontier devices $a_\ell(j)$ but differ in the finish times $L_j^\ell$ and in the devices to which their outputs have already been transferred to.
Assigning $s_i$ to an admissible device extends a label as
\begin{equation}
L_i(\ell,d) = T_{\mathrm{comp}}(s_i,d)
+ A_i\!\left(d,\{(a_\ell(j),L_j^\ell)\}_{j\in\mathrm{pred}(i)}\right),
\label{eq:dp-transition}
\end{equation}
where $A_i$ follows Eqs.~\ref{eq:arrival-link}--\ref{eq:arrival} with $Q_{d'\to d} =A_{i-1}$ when $a(i-1)=d$, and 0 otherwise.

In the fan-in example above, a frontier assignment fixes the devices of $s_{i-2}$ and $s_{i-1}$, and each label under it stores their finish times.
Placing $s_i$ on $d$ combines these finish times through Eq.~\ref{eq:dp-transition}, and the new label stores $L_i(\ell,d)$ together with the finish times of the stages that remain in the frontier.
The number of nodes that enter the frontier before being resolved is bounded by the width ($\omega$) of the residual connections in the graph (how many stages are skipped), so the number of tracked states due to the frontier is at most $|\mathcal{D}|^{\omega}$.
\subsubsection{Memory as a resource dimension}
The frontier-state DP makes the latency of a placement computable but does not ensure that it fits in memory.
HELIX extends the DP state with an occupancy vector $u$, similar to solutions to the multi-dimensional multiple-choice knapsack problem~\cite{MMKP2004}, whose entries carry the aggregate footprint on each device.
Extending a partial placement $\ell$ with $s_i\mapsto d$ updates only that device's entry, $u^{\ell\oplus(s_i\mapsto d)}_{d} = u^{\ell}_{d} + \lceil m(s_i,d)/b\rceil$, and partitions where $u^{\ell'}_{d} > M_d$ are discarded.

However, not every device needs an entry in the occupancy vector for all serving regimes and modes.
We consider $\mathcal{T}$ the set of memory-tracked devices, which contains only the \textit{overflowable} ones, i.e., those whose capacity would be exceeded if every stage listing $d$ among its candidates $C_i$ were assigned to it.
Therefore, the tracked occupancy vectors are limited to devices $d\in\mathcal{T}$.
A GPU, for instance, supports every operator and becomes overflowable once parameters and KV cache/SSM exceed its capacity.
Reducing the memory-tracked devices is crucial as search time grows quickly with $\mathcal{T}$.

Adding occupancy to the state enlarges the search space, as partial placements that would be interchangeable under latency considerations, now can differ in how they occupy memory.
To keep the states tractable, HELIX prunes states by Pareto dominance.
A label is dominated by another if the latter has no greater running latency, no later finish time for every live frontier stage, and no greater tracked occupancy.
The effectiveness of pruning in keeping the search time small depends on how often occupancies are comparable, so byte-level precision makes cases where dominance occurs rare, keeping apart states whose occupancies differ too little to affect any later feasibility decision.
Therefore, occupancy is tracked in integer bucket units of size $b$.
Each stage occupies then $\lceil m(s_i)/b \rceil$ and each device has capacity $\lfloor M_d/b \rfloor$.
Further implementation details are in Appendix~\ref{app:algorithm}.
\subsection{Complexity and optimality}
\paragraph{Optimality:} DP finds an optimal placement for the reduced-graph objective in Equation \ref{eq:objective}.  Pareto pruning discards a partial-placement label only if another label with the same frontier label has lower or equal running latency, finish time on every live frontier stage, and occupancy on every device.
We treat pruning of the memory representation as a heuristic to reduce the amount of tracked states and evaluate the resulting placement quality empirically.
Additionally, accounting for memory occupancy in buckets may be a source of approximation.
A larger $b$ leaves fewer possible occupancy values per device, so more memory states match and get pruned, reducing the search space and time.
However, faster search time comes at the cost of accuracy in how the occupancy is being tracked, as coarser buckets can conservatively reject feasible placements (see Appendix Figure \ref{app_fig:mem_ablation}a).
We set $b$ to the greatest common divisor of all stage footprints. Capacity rounding is exact since all footprints are multiples of $b$. 

\paragraph{Complexity:} Without memory overflow, search over the DAG visits each stage just once, in addition to extending each frontier state over admissible devices.
With $\omega$ stages in the frontier, there are at most $|\mathcal{D}|^{\omega}$ frontier assignments, resulting in $O(N\,|\mathcal{D}|^{\omega+1})$ transitions when one sufficient state is kept per assignment, against the $O(|\mathcal{D}|^{N})$ complexity of exhaustive search.
In transformer-like models, $\omega$ does not depend on the model's depth and usually stays between 2 and 4 stages, so $|\mathcal{D}|^{\omega}$ stays small and the search is linear in $N\cdot D$.
Instead, tracking memory increases the amount of labels retained per stage.
Considering $P$ as the maximum number of nondominated labels retained per key.
Labels sharing a frontier key and an occupancy vector are compared on arrival time alone, so at most one is kept per occupancy vector, giving $P \le \prod_{d\in\mathcal{T}}(\lfloor M_d/b\rfloor+1)$, where $\lfloor M_d/b\rfloor$ refers to the amount of values each occupancy vector can take, which depends jointly on the memory capacity and the bucket size.
The search therefore considers $O(NP|\mathcal{D}|^{\omega+1})$ label extensions, with additional cost for arrival evaluation and dominance checks. 
$P$ grows polynomially with $N$ for a fixed $\mathcal{T}$ until it saturates the memory constraint, and then becomes linear in $N$, which can happen in early stages of the graph under high memory constraints.
In practice, HELIX targets regimes where each accelerator hosts most of the parameters of its supported operators, since heterogeneous platforms pair devices to take advantage of their compute properties rather than memory aggregation, so $\mathcal{T}$ stays small. 
If model placement is dictated by memory constraints instead of complementary hardware strengths, search should lean towards resolving sharding configuration as well.

\section{HELIX: the execution path in vLLM}\label{sec:vllm_HELIX}
\begin{figure}[ht]
    \centering
    \includegraphics[width=0.475\textwidth]{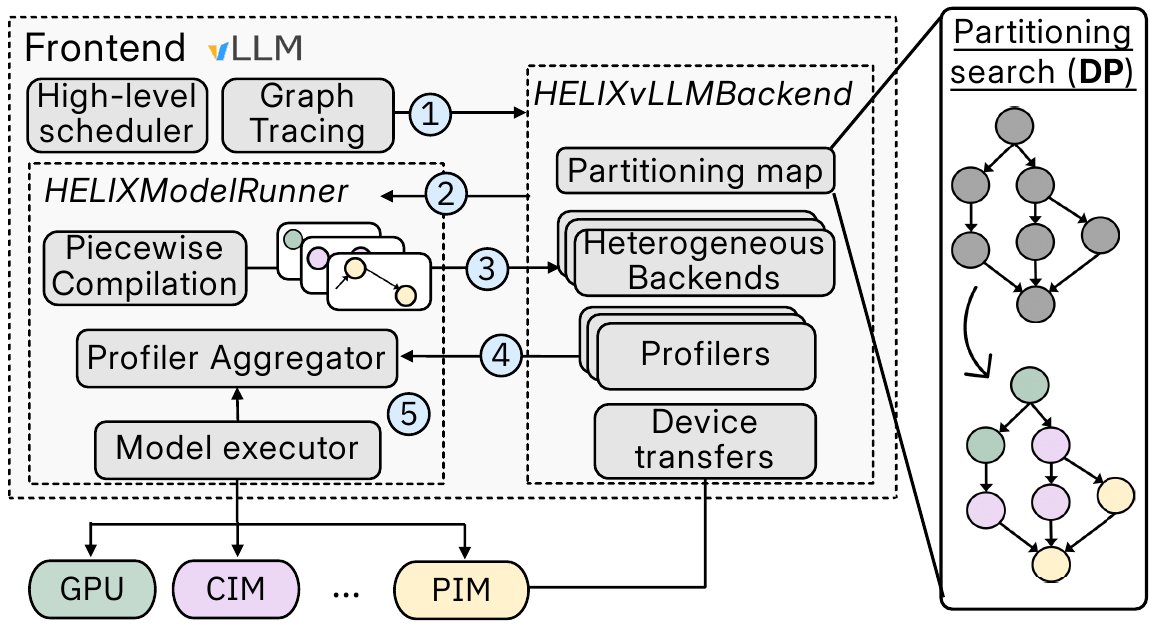}
   \vspace*{-5pt}
    \caption{HELIX software stack: graph partitioning based on DP \circled{1}, piecewise compilation \circled{2} on heterogeneous backends \circled{3}, and profiler feedback \circled{4} through execution \circled{5}.}
    \label{fig:architecure}
        \vspace{-5pt}
\end{figure}

The placement resolved in the previous section defines which device each stage belongs to, and leaves open the mechanism by which each fragment is compiled for its target device, and how activations cross between them.
Figure~\ref{fig:architecure} illustrates the HELIX software stack.
At the frontend, HELIX integrates with vLLM, inheriting its scheduling and KV-cache management.
Model loading is intercepted by \texttt{HELIXvLLMBackend}, which hooks into TorchDynamo's tracing pipeline and splits the traced graph for a given workload (derived from the model's context length and vLLM's maximum number of concurrent sequences), model and available hardware, resolving operator-to-backend assignments from kernel support, profiled runtime cost and memory capacity~\circled{1} with the methodology described in \ref{ref:dp}.
The \texttt{HELIXModelRunner} reuses the existing \emph{piecewise compilation} which is used to compile CUDAGraph safe and unsafe regions, to convert each subgraph into a standalone artifact specialized to its target and stitching them with designated transfer modules~\circled{2}.
Compute artifacts are dispatched sequentially, while transfer modules are issued asynchronously so that a transfer starts as soon as its producer finishes and overlaps with compute on other devices, as assumed by the cost model (Eqs.~\ref{eq:arrival-link}--\ref{eq:arrival})~\circled{3}.
Each hardware is exposed to the stack through a vLLM-compatible compilation interface, so that graph tracing can verify operator support.
Adding new hardware only requires implementing this interface and profiling the costs used for partitioning, so a vendor can evaluate an emerging accelerator within a full serving stack without porting an entire model.

Compute costs are collected per device from either hardware profilers on physical accelerators or software simulators modeling emerging hardware, allowing HELIX to mix both sources within a single deployment.
These costs are then aggregated in the model runner to report user metrics~\circled{4} during execution~\circled{5}.

\subsection{Interception point: piecewise compilation}\label{sub:tracing}
vLLM V1 integrates \texttt{torch.compile} as a core pipeline component: \texttt{TorchDynamo} captures tensor operations into a computation graph, and \texttt{TorchInductor} lowers it to kernels.
Since some operations are size-variable (e.g., attention over a KV cache of unpredictable length), vLLM uses \emph{piecewise CUDA Graphs}, splitting the graph into CUDA-Graph-safe and -unsafe regions executed as independently dispatchable artifacts, with attention detached from the rest.
HELIX intercepts at this attention/non-attention boundary and extends it, partitioning the graph at any boundary where target hardware changes and turns each piece into hardware-specific subgraphs, leaving vLLM's GPU-side optimizations (kernel fusion, CUDA Graph replay, shape specialization) untouched.
vLLM suits HELIX because its graph tracing is device-independent, with boundaries defined over generic tensor operations, so any \texttt{torch.compile}-compatible accelerator can become a dispatch target.
Engines like TensorRT-LLM~\cite{tensorrt_llm} or SGLang~\cite{sglang}, whose segmentation is bound to the GPU execution model,  lack a comparable interception point for non-CUDA substrates.
\subsection{Heterogeneous graph partitioning and dispatch}
The resolved placement from Section~\ref{ref:dp} enters the stack here.
HELIX intercepts the pre-dispatch \texttt{torch.fx.Graph}, traverses it node by node, and assigns each node to the backend that the DP partition places its DAG stage to.
Contiguous runs of nodes sharing a target backend are then fused into maximal subgraphs, minimizing backend transitions and preserving kernel-fusion opportunities within each partition.
 
At every boundary where the target backend changes, HELIX inserts a dedicated transfer module, resolved by data flow rather than by the positional adjacency of subgraphs, so that transfers carry only the tensors the target backend actually consumes.
Registering the module inserts a node into the stitching graph, redirects all affected consumers to the transferred copies and re-compiles.
Each transfer becomes its own subgraph in the dispatch list, interleaved with the compute subgraphs.
This insertion point may also serve as the integration surface for communication backend development (such as NIXL~\cite{ovidiusm2026ai}) as heterogeneous platform support matures.
Each resulting subgraph is then compiled by its respective backend into standalone compiled artifacts and dispatched independently at run time.

\section{Evaluation Methodology}
\subsection{Models and Metrics}
\textbf{Llama3.1-8B} is a dense transformer model, while \textbf{Mixtral-8x7B} replaces the FFN with a sparse mixture of experts, both with quadratic grouped-query attention (GQA). \textbf{Bamba-9B} is a hybrid that interleaves a few GQA layers among state-space mixer layers, exhibiting two profiles on the weightless computations.
Because partitioning decisions are driven by the mix of computational primitives rather than by any specific release, each model stands for its architectural family.
Newer models within these families largely reuse the same primitives, varying mainly the attention mechanism and proportions between dynamic and static operators, while producing DAGs of comparable size (Appendix Table~\ref{tab:nodes_vs_stages}).

We focus on the decode phase, which dominates serving time and is where heterogeneous accelerators target memory-bound operators; prefill can be handled by existing P/D disaggregation, although the partitioning search applies to it unchanged with prefill-shaped profiles.
Partitions are judged by the latency a single decoding step and the generation throughput over multiple steps.

\subsection{Hardware platforms and profiling methodology}
\textbf{GPU (H100-80GB SXM, B200-192GB SXM).} Latencies come primarily from the public \textit{PerfDatabase}~\cite{aiconfigurator}, which reports single-kernel latencies profiled on vLLM's execution path.
We follow its profiling protocol\footnote{\url{https://github.com/ai-dynamo/aiconfigurator}} to add in-house measurements for ISLs and batch sizes it does not cover (GEMMs, GQA) and for the Mamba mixer, validating our data against the released data where they overlap.

\textbf{DRAM-PIM.} We adopt the AttAcc~\cite{Park2024AttAcc} accelerator, built on 8-Hi HBM3 stacks, which targets attention.
In batched decoding, the score ($QK^\top$) and context ($\alpha V$) operations reduce to per-request GEMVs over the KV cache.
AttAcc places GEMV units at the bank level so that they execute at DRAM internal bandwidth rather than over the external HBM I/O.
We instantiate a single-GPU-scale system of 5 AttAcc stacks (equivalent to the HBM of an A100-80GB SXM) and obtain attention latencies from the authors' open-source simulator\footnote{\url{https://github.com/scale-snu/attacc_simulator}}.

\textbf{3D compute-in-memory (3DCIM).} 3DCIM is an analog in-memory computing accelerator built around vertically stacked non-volatile memory, which maps linear layers to tier-level matrix--vector multiplications executed within the array~\cite{ICCD20253dnand, ISCA263dnandMoE, Bchel2025Efficient}.
Stationary weights eliminate parameter-fetch traffic, making it well suited to compute-bound fully connected and MoE expert layers.
We model an accelerator inspired by \cite{Bchel2025Efficient} with a capacity of $\sim$ 86B parameters, and a 200 ns MVM integration time.
The 3DCIM simulator serves as a backend for HELIX-vLLM execution, serving to demonstrate how a simulator and hardware profiling backends can be implemented simultaneously in HELIX-vLLM. Further details on the 3DCIM system simulator are available in Appendix \ref{3dcim_simulator}.

Finally, connections between accelerators are assumed peer-to-peer using NVLink 3.0 with 300 GB/s and a startup latency of 2 microseconds \cite{NVLinkStartup}.
\subsection{Baselines}
We test three heterogeneous system configurations: \textbf{3DCIM + GPU}, \textbf{AttAcc + GPU}, and \textbf{3DCIM + GPU + AttAcc}.
The pairwise configurations reflect the GPU-plus-accelerator pairings discussed in Section~\ref{sec:background}, where a fixed partitioning strategy is prescribed by the design itself.
Concretely, for 3DCIM + GPU, the baseline partition follows mainstream A/F as proposed in \cite{nvidiaInsideNVIDIA, Zhu2025MegaScale} where either the FFN block or the experts are offloaded from the GPU.
For DRAM-PIM + GPU, the baseline follows the strategy adopted in AttAcc~\cite{Park2024AttAcc} where only the attention is offloaded from the GPU; this is referred to in the rest of the paper as the \textit{AttAcc} baseline.
All methods use identical profiles, operator-support constraints, communication assumptions and memory budgets.
\section{Results}
\subsection{DP-based Partitioning search}
We first assess the efficiency and correctness of HELIX’s DP-based partitioning search. Table~\ref{tab:search} shows execution times for HELIX DP-based partitioning search against an exhaustive-like baseline over DAGS of different structures. The first four rows show search times for a small synthetic DAG of 16 stages isolating the effects that add complexity to HELIX DP. The first two rows show the effect of the frontier, raising slightly the search time from 1.4 milliseconds to 3 milliseconds when adding residual skips of width 3 every four stages (similar to the pattern in transformers). The third and fourth rows isolate the effect of tracking memory across the search for three equally-constrained devices in terms of capacity where $\sum m(s_i) \approx 1.4\times \sum M_d$, and shows how state pruning, based on Pareto dominance, keeps the search time short (785 ms vs. 267 s) without compromising the final result. On all synthetic DAGs, the DP search returns the same solution as the exhaustive search.
For a realistic graph, enumerating the entire search space is intractable for $3^{162}$ combinations, so we instead conducted Monte Carlo simulations, sampling 4,000 initial random placements and timed them out after 30 minutes of search, and comparing their cost against the DP solution. None of the random searches returned a partition with lower latency than DP. Further ablations over the algorithms can be found in Appx. Section \ref{app:algorithm}. Placing the full set of stages of Llama takes roughly 16 milliseconds, which makes the partition search seamless at compilation time. 
\begin{table}[h!]\label{tab:search}
\centering
\renewcommand{\arraystretch}{0.9}
\begin{tabularx}{\linewidth}{c|c|c|c|>{\centering\arraybackslash}X}
\toprule
\multirow{2}{*}{\textbf{DAG}} & \multirow{2}{*}{\textbf{$\mathcal{D}$/$\mathcal{T}$}} & \multirow{2}{*}{\textbf{HELIX}} & \textbf{Baseline} & \multirow{2}{=}{\textbf{Est. \newline Latency}} \\ \cline{4-4}
 &  &  & \textbf{Exh.} &  \\ \toprule
Chain(16) & 3/0 & 1.4 ms & 3342.6 s & = \\ \hline
Skip(16) & 3/0 & 3.1 ms & 3129.8s & = \\ \hline
\multirow{2}{*}{Skip(16)} & 3/3 & 267 s\textsuperscript{*}& \multirow{2}{*}{2018.9} & = \\ \cline{2-3} \cline{5-5} 
 & 3/3 & 785 ms &  & = \\ \midrule
\multicolumn{1}{l|}{\multirow{2}{*}{\begin{tabular}[c]{@{}l@{}}Llama 8B\\  (162)\end{tabular}}} & \multicolumn{1}{l|}{\multirow{2}{*}{3/0}} & \multicolumn{1}{l|}{\multirow{2}{*}{15.967 ms}} & \textbf{Rand.} & \multicolumn{1}{>{\raggedright\arraybackslash}X}{\multirow{2}{=}{3.1$\pm$0.45$\times$}} \\ \cline{4-4}
\multicolumn{1}{l|}{} & \multicolumn{1}{l|}{} & \multicolumn{1}{l|}{} & $<$ 30 min & \multicolumn{1}{>{\raggedright\arraybackslash}X}{}
\\
\bottomrule
\end{tabularx}
\vspace{-5pt}
\caption{HELIX (DP) search time is compared against exhaustive search on small graphs (top four rows; \textsuperscript{*} = DP without state pruning), with the last column showing whether estimated latency matches (=). On the full LLM graph (bottom row), it is compared against the average of 4,000 random searches (30-minute timeout), whose found partitions have higher latency than DP’s.}
\label{tab:my-table}
\vspace{-5pt}
\end{table}
\newline
Having validated the DP-based search, we now apply it to different heterogeneous systems to show the advantage of tuned partitions over standard disaggregation techniques when different accelerators are combined. Subsections \ref{sub:afd}--\ref{sub:3way} each target a different heterogeneous platforms, resolving the partitions and reporting the decode latency the search estimates from pre-profiled data across a range of models, query counts, and context lengths.
\subsection{A/F-style system: GPU + 3DCIM}\label{sub:afd}
\begin{figure}[ht]
    \centering
\includegraphics[width=0.495\textwidth]{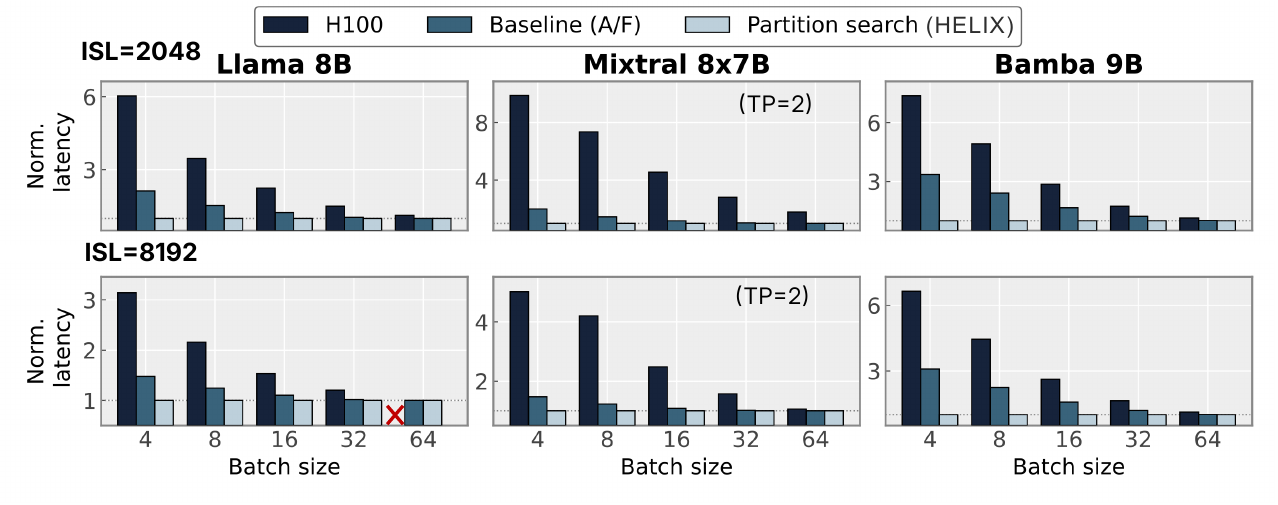}
\vspace{-20pt}
    \caption{Single-step decoding latency of the H100-only and A/F baselines, normalized to HELIX. For Mixtral 8x7B, which does not fit on a signle H100, the baseline uses tensor parallelism (TP) = 2. Partitions can be found in Appendix Figure \ref{app_fig:partitions_vs_afd}.}
    \label{fig:afd_system}
    \vspace{10pt}
\end{figure}

Here we test a heterogeneous system composing an H100 and 3DCIM accelerator, which maps directly onto the A/F disaggregation trend and therefore is the selected baseline.
Figure~\ref{fig:afd_system} reports the latency of a single decoding step for three strategies: H100-only execution, the standard A/F split (FFN or expert layers on 3DCIM, all other operations on GPU), and the optimal partition found by the DP search, to which the other two are normalized. 
At small batch sizes, where decoding is dominated by weight fetching, the HELIX partition is $\sim$3.14--6$\times$ faster than single-H100 for Llama 8B and $\sim$4.48--7.3$\times$ faster for Bamba 9B, for ISL of 8K and 2K, respectively.
For Mixtral 8x7B, whose parameters do not fit on a single GPU, the baseline is 2xH100, and HELIX is $\sim$5--9.8$\times$ faster over the same ISL range.
Pairing a GPU with CIM core that eliminates memory-bottleneck is already advantageous, and offloading only the FFN/MoE (A/F) captures much of the gain, though it is not the optimal split in all scenarios.
With up to 32 concurrent queries the optimal partition differs the most from A/F, especially for Bamba 9B.
Bamba’s hybrid structure makes standard A/F hard to apply, and our search discovers the better split automatically by assigning all GEMMs to 3DCIM, including those around the mixer and attention, rather than only the FFN. 
Its mixer projections have larger input and output dimensions than the attention projections in Llama 8B and Mixtral 8x7B and account for most of its execution time (see Appx. Figure \ref{app_fig:exec_time_per_model}), so offloading them yields speedups of roughly 1.7--3.2$\times$ over A/F depending on batch size. These gains roughly hold at longer contexts, since the SSM state update, unlike attention, does not grow with context length\footnote{Bamba-9B-v1 has a maximum model length of 4096 tokens, so longer contexts are shown to be representative of how a model less bound by attention would perform at higher context lengths.}.
For Llama 8B and Mixtral 8x7B, KV-cache fetching increasingly dominates decode latency as context grows, so the optimal partition lands gains only ~2.2--1.1$\times$ over A/F for batch sizes up to 16 and context lengths of 2K to 8K, respectively.
Around batch size 64, the HELIX partition converges to A/F, meaning the optimal split offloads only the FFN/MoE to the 3DCIM accelerator.
Beyond this batch count, the best partition assigns all GEMMS back to the GPU for all models except for the Mixtral, because transfer latencies increase with batch size and mask the per-GEMM speedups 3DCIM can have over H100. 
For Mixtral, the experts remain in 3DCIM since loading all eight experts’ weights keeps its computation memory-bound even at higher concurrency. 
This results in $\times$1 w.r.t. GPU-only baseline, but for this serving scenario, attention dominates execution, so all partitions perform similarly.
\subsection{GPU + DRAM-PIM}\label{sub:attacc}
The limited gains of any HELIX partition at large batch sizes and context lengths stem from attention remaining resident on the GPU.
We now relax this constraint and study a system where the second accelerator can also execute attention (i.e. DRAM-PIM), allowing the partition search to split attention as well as the projections across both devices.
For DRAM-PIM, GEMM latency is estimated with a roofline model.
Mixtral 8x7B is excluded from this study because mapping its experts to DRAM-PIM is non-trivial, and its expert weights ($\sim$90,GB in FP16) exceed the GPU’s 80,GB memory.
 \begin{figure}[ht]
    \centering
\includegraphics[width=0.475
\textwidth]{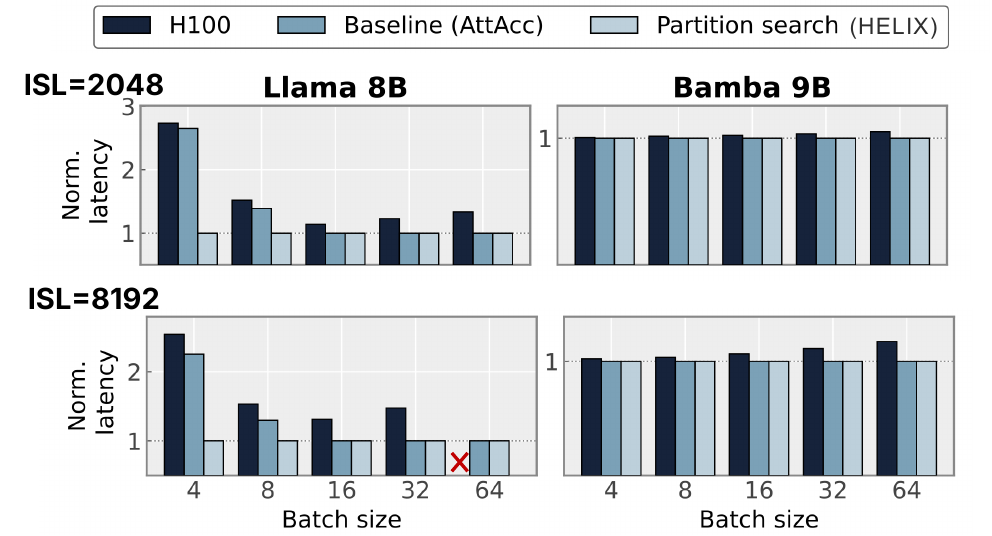}
\vspace{-10pt}
    \caption{Single-decoding-step output token latency of the H100-only and \textit{AttAcc}-style disaggregation baselines, both normalized to the latency given by the HELIX partition.}
    \label{fig:attacc_system}
    \vspace{-10pt}
\end{figure}
The \textit{AttAcc} simulator models multi-head attention (MHA) instead of GQA, but since the authors report no latency benefit of GQA on PIM~\cite{Park2024AttAcc}, we consider the simulated MHA latencies with a GQA memory footprint.
Figure~\ref{fig:attacc_system} compares GPU-only and \textit{AttAcc}-style disaggregation against the HELIX partition.
For 4 queries, HELIX finds partitions that place both GEMMs and attention on DRAM-PIM for Llama 8B, achieving 2--2.7$\times$ speedups over \textit{AttAcc}.
However, GEMMs on DRAM-PIM are beneficial only at low batch sizes as its compute throughput ($\sim$30 TFLOPS) is far below the H100's.
Beyond 16 queries, \textit{AttAcc}-style placement is the one found to be optimal.
Notably, at 64 queries with 8K tokens in context, memory pressure hits, and execution under standard \textit{AttAcc} disaggregation or GPU-only would be infeasible.
HELIX still finds a feasible partition by moving part of the attention heads from DRAM-PIM to H100, which now would host the GEMMs along with 13 attention layers.
Instead, the chosen partitioning for Bamba 9B is exactly \textit{AttAcc}-style, as the large input and output projections are faster on the H100 regardless of the batch count. Detailed partitions for Llama 8B are available in Appx. Figure \ref{app_fig:partitions_vs_attacc1}.
\begin{figure*}[t!]
    \centering
\includegraphics[width=0.975\textwidth]{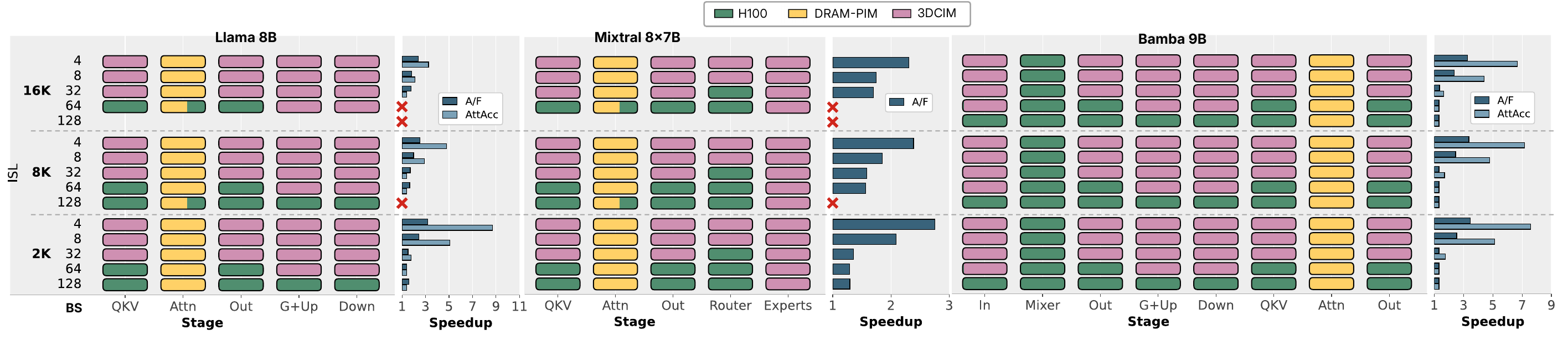}
\vspace{-10pt}
    \caption{Partitions found for the three-way system (H100 + 3DCIM + DRAM-PIM) across batch sizes and sequence lengths, and the single-decoding-step speedup over the A/F and \textit{AttAcc}-style baselines. \textcolor{red}{$\times$} marks memory-infeasible configurations.}
    \label{fig:3way_h100}
    \vspace{-10pt}
\end{figure*}
Therefore, gains over GPU-only execution shrink across all workloads for Bamba 9B, since only 3 of its 32 layers contain GQA and the rest are SSM blocks. 
Models like Bamba do not benefit from \textit{AttAcc}-style partitions. Prior work has explored accelerating SSM blocks with PIM architectures \cite{PIMBA}, which we leave for future study. 
Yet, the previous results in \ref{sub:afd} showed that HELIX still finds partitions that outperform more suitable baselines such as A/F for this model.
This highlights how the best strategy depends jointly on the model architecture and the system configuration.
With newer models adopting different attention variants (e.g., multi-head latent attention, SSM, gated delta attention) and different ratios of dynamic to static operations, a strategy designed for one model family, such as \textit{AttAcc}, does not directly transfer to another.
\subsection{Three-way heterogeneity}\label{sub:3way}
\begin{figure}[!ht]
    \centering
\includegraphics[width=0.495\textwidth]{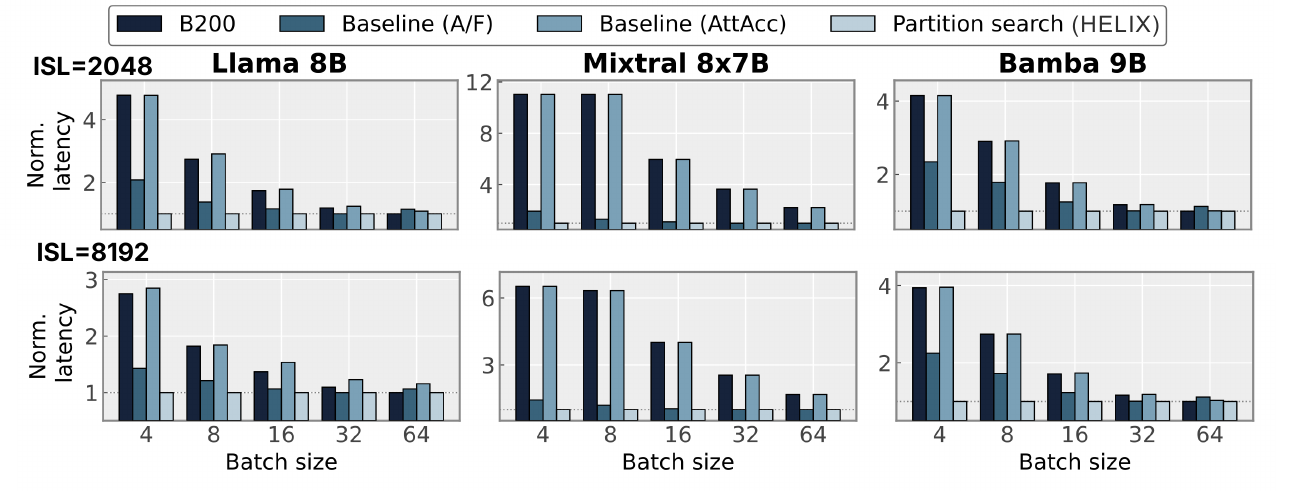}
\vspace{-15pt}
    \caption{Single-decoding-step latency for B200 + 3DCIM + DRAM-PIM and speedup over pair-wise baselines. Partitions in Appendix Figure~\ref{fig:appx_3way_b200}}
    \label{fig:3way_b200}
        \vspace{-12pt}
\end{figure}

In this section, we study a more complex system of three hardware components to illustrate how HELIX can guide heterogeneous system design by simulating as-yet unrealized hardware combinations on a given model and workload.
The two pairwise studies \ref{sub:afd} and \ref{sub:attacc} show that 3DCIM is a strong  accelerator for most of the GEMMs, moving towards H100 for larger batch count as computation gains are masked by transfer latencies, while DRAM-PIM remains the fastest attention substrate across all evaluated workloads.
However, no existing disaggregation strategy covers how to partition a model across three components with such different properties, as both A/F and \textit{AttAcc}, are GPU-centered and only serve as pairwise references.
Figure~\ref{fig:3way_h100} shows the partitions found for each workload and their single decoding-step speedup achieved over the two baselines.
Below 64 concurrent queries, all projections stay on 3DCIM, and attention is consistently assigned to DRAM-PIM, a placement neither reference or existing system in literature can express.
In this range, the optimal partition is 1.8--3.5$\times$ faster than A/F and 1--8.7$\times$ faster than \textit{AttAcc} for Llama 8B, and 1.8--3.6$\times$ faster than A/F for Mixtral 8x7B depending on batch size and context.
Notably, neither baseline is optimal across the full range of workloads, since A/F is the better option below 16 concurrent queries at every evaluated context length and  the \textit{AttAcc} one beyond that point, while HELIX partitions outperform both in all cases.
The GPU could still be involved in running operations such as Rotary Position Embedding (RoPE).
Since our DAG abstraction absorbs RoPE into the attention stage, we assume it can run on the DRAM-PIM base die alongside Softmax, avoiding costly data transfers. 
Bamba again shows the largest gains over A/F. 
Because its Mamba mixer is tied to the GPU, every partition spans all three accelerators, which opens opportunities for finer-grained pipelined scheduling as future work, using sub-batching to increase resource utilization.
\vspace{-5pt}
\paragraph{Replacing the H100 with a B200}
To examine whether these placements persist when the GPU changes, Figure~\ref{fig:3way_b200} repeats the study with a B200 replacing the H100.
The B200's higher compute throughput and memory bandwidth makes it more competitive for both GEMMs and attention, shifting where each is best placed.
At batch size 4 and context length 2K tokens, for instance, HELIX partition offloads all GEMMS to 3DCIM rendering a  1.9$\times$ speedup against A/F, compared to 3$\times$ speedup over A/F when applying the same partition with the H100-based three-way system, since now GEMMs in the B200 are slower than 3DCIM but faster than H100.
As batch size grows, partitions for Llama 8B and Bamba 9B converge to A/F at 32 queries (instead of 64), and, to B200-only execution at 64.
The B200 matches and starts to outperforms DRAM-PIM for attention execution around batch count 16.
This suggests that B200’s higher memory bandwidth based on HBM3e, combined with its compute cores, is enough to compete with the benefits of near-memory computing on an HBM3 memory system.
Changing only the GPU scale shifts partition boundaries, confirming that each hardware combination demands its own partition search rather than a fixed strategy.
\subsection{Offline serving with HELIX-vLLM}\label{sub:vllm}
\begin{figure}[ht]
    \centering
\includegraphics[width=0.45\textwidth]{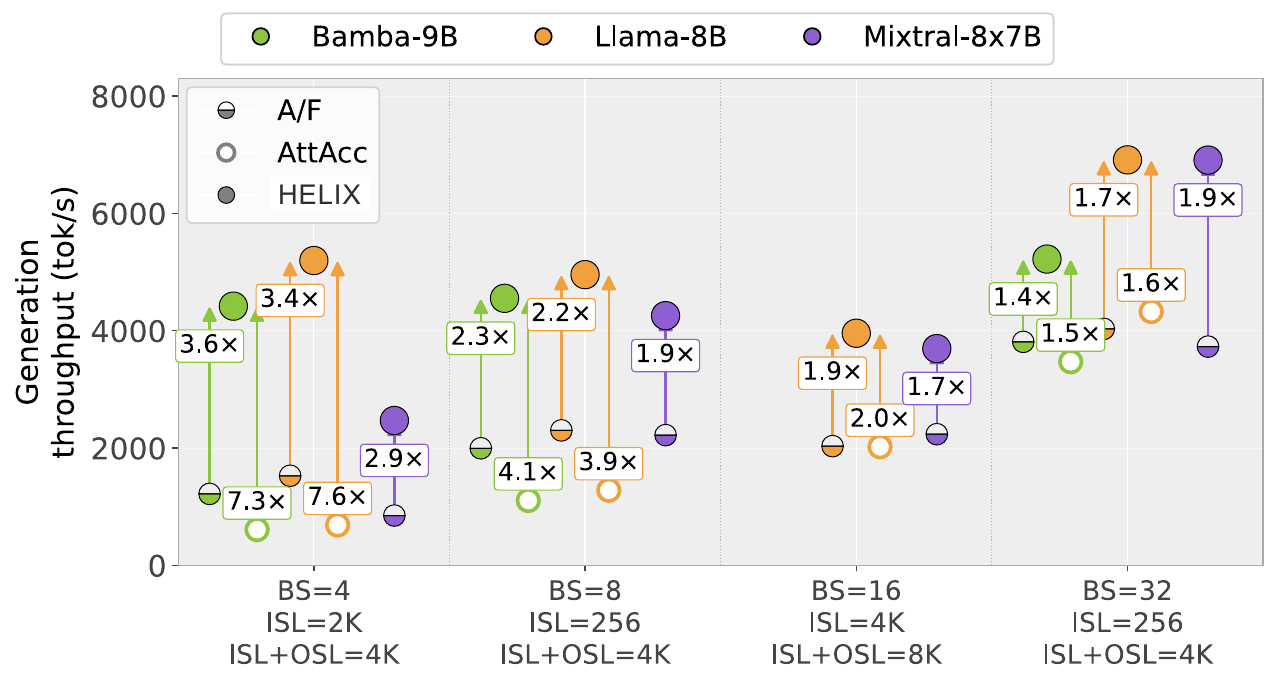}
\vspace{-7.5pt}
    \caption{Generation throughput over complete decoding graphs in HELIX-vLLM for the three-way heterogeneous partitions found by HELIX, compared to the A/F and \textit{AttAcc}-style baselines.}
    \label{fig:e2e_throughput}
    \vspace{-12pt}
\end{figure}

Figure~\ref{fig:e2e_throughput} reports end-to-end generation throughput with the found partitions deployed in vLLM over multiple decoding steps.
Unlike the single-step estimates above, which add up the costs of the profiled primitives only, these runs account for every operation in the parent graph, including element-wise and normalization operations, except for RoPE when executed in DRAM-PIM.
\section{Related Work}
Device placement of DNN graphs on distributed environments has been widely studied.
Tarnawski et al.~\cite{NeurIPS2020DNNDAG} formulate operator placement across identical accelerators and CPUs as an integer program for latency and a dynamic program for pipelined throughput, where split decisions are triggered by memory capacity and graph parallelism.
PipeEdge~\cite{PipeEdgeDP} splits chain-like DNN graphs at the layer granularity across heterogeneous edge devices to minimize the slowest pipeline stage within memory constraints. 
HELIX instead targets residual LLM graphs with hundreds of stages across multiple accelerator types, where splits are driven by per-operator performance rather than capacity alone, and resolves placement within compilation-time budgets.
For LLM inference specifically, methods in literature have started to move beyond fixed A/F splits towards finer-grained  execution. PAPI~\cite{PAPI2025asplos} and DOPS~\cite{DOPS} co-design runtime operator scheduling with tailored hardware to dispatch kernels between DRAM-PIM and a GPU or an NPU, respectively, deciding among predefined kernel classes for a fixed hardware pairing.
In contrast, HELIX is hardware-agnostic, searching placements over the full model graph for arbitrary accelerator combinations within a serving engine.
OpWeave~\cite{Opweave} also argues A/F boundaries are suboptimal for newer models and proposes general operator partitions within vLLM to minimize the dollar cost per million tokens. Its heterogeneity is limited to GPUs of different capability rather than accelerators with complementary strengths where each excells different operators and serving regimes. The two approaches are therefore orthogonal, addressing the same motivation under different objectives and hardware assumptions.
\section{Conclusion}
\label{sec:conclusion}
HELIX proposes method to resolve workload-specific operator partitions and dispatch them across heterogeneous hardware within a unified vLLM engine.
By treating hardware as graph executors, we enable emerging accelerators (e.g., PIM, 3DCIM) with partial operator support to participate in LLM serving. 
Future work can extend HELIX to optimize energy consumption and resource utilization alongside latency, by targeting pipelined execution when multiple accelerators are available. 
Pre-compiling partitions for several serving regimes would further enable online adaptation as well as auto-tuning through runtime profiling to inform partitioning decisions.
\section{Acknowledgments}
We thank Thomas Parnell, Alberto Mannari, Garima Singh, and Irem Boybat for their valuable technical discussions. 
We also acknowledge Vijay Narayanan and Jeff Burns for their managerial support.

\newpage

\bibliography{example_paper.bib}
\bibliographystyle{template}

\section{AI Usage Disclosure}
Generative AI was used to assist with code development, including plotting code for visualizing our experimental results, and with grammatical and typographical proofreading, not for writing.

\newpage
\clearpage
\appendix
\section{Validation of the Cost-Model Assumptions}
Our placement problem formulation relies on three assumptions: (i) end-to-end latency is dominated by a small set of computational primitives, (ii) a roofline-based cost model does not give a sufficiently accurate latency estimate for these primitives, requiring per kernel profiling, and (iii) kernels profiled in isolation are representative of their behavior within full model execution. This section provides empirical evidence for each.

\begin{figure}[ht]
    \centering
    \includegraphics[width=0.45\textwidth]{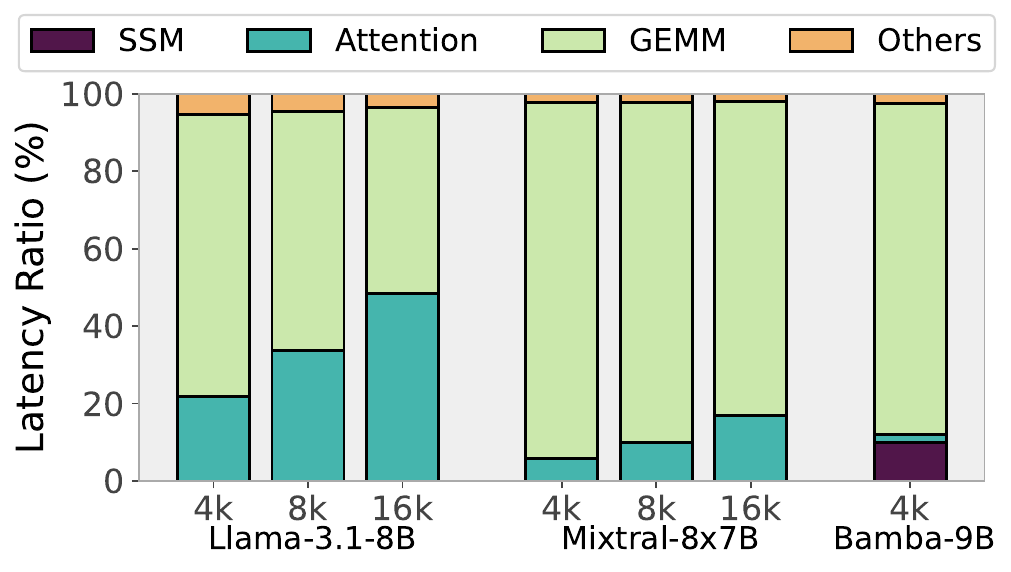}
    \vspace{-10pt}
    \caption{Breakdown of end-to-end execution time on an A100 between the main computational primitives (GEMMs, attention, state scan and update) and the remaining operations.}
    \label{app_fig:exec_time_per_model}
\end{figure}

\begin{figure}[ht]
    \centering
    \includegraphics[width=0.495\textwidth]{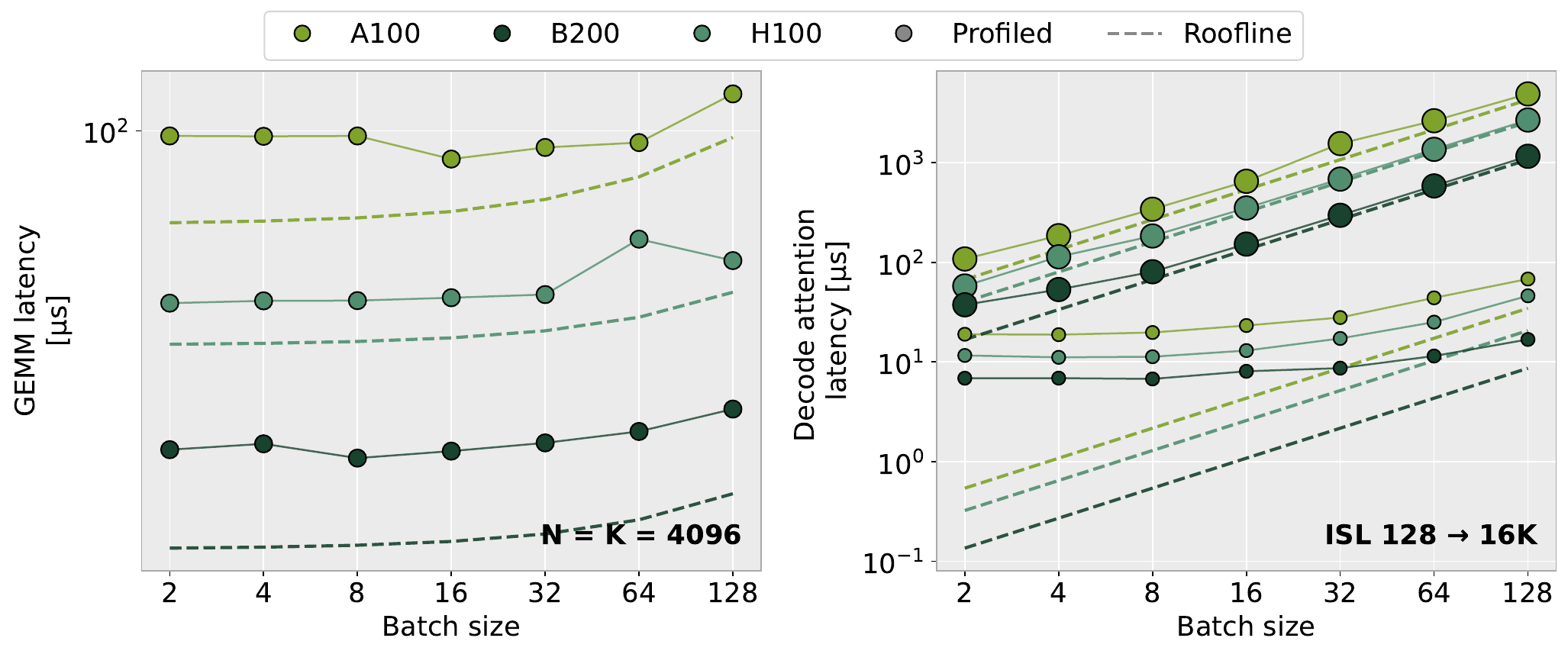}
    \vspace{-10pt}
    \caption{Latency estimation based on roofline cost-model vs. obtained from profiling for A100, H100 and B200 GPUs.}
    \label{app_fig:roofline_vs_profile}
\end{figure}

\begin{figure}[ht]
    \centering
    \includegraphics[width=0.5\textwidth]{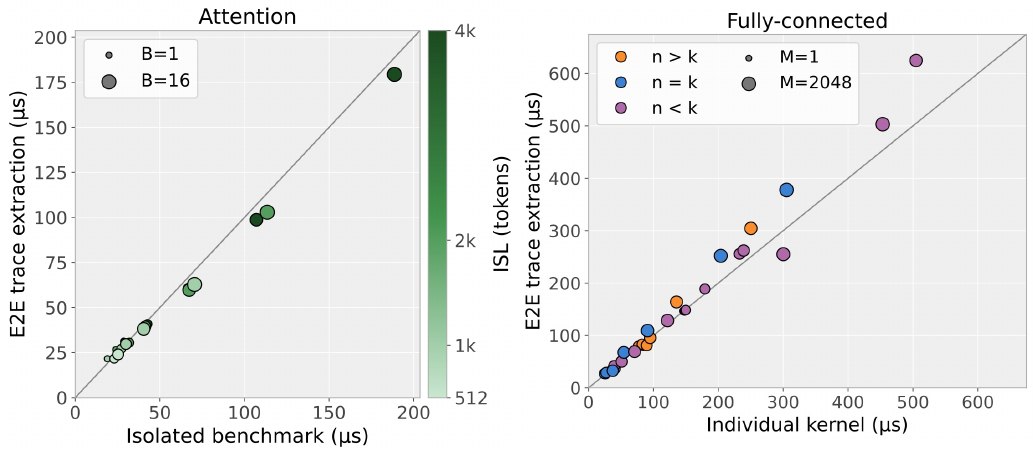}
    \caption{Individual kernel profiling versus end-to-end latency kernel extraction}
    \label{app:fusion_gpu}
\end{figure}

\paragraph{Dominance of the main computational primitives on the execution time:}
Figure~\ref{app_fig:exec_time_per_model} breaks down end-to-end execution time on an A100 GPU. GEMMs, attention, and state scan/update operations account for over  [95\%] of the total runtime across all evaluated models, while the remaining operations (e.g., normalization, activations, element-wise ops) contribute only in a small fraction.
\paragraph{Accuracy of roofline cost models:}
Figure~\ref{app_fig:roofline_vs_profile} compares the latencies predicted by the roofline model with those measured through profiling on A100, H100, and B200 GPUs. For GEMMs, the estimates only loosely track the measurements, since achieved resource utilization varies across operation shapes. In contrast, strictly memory-bound attention (ISL of 16K tokens, shown as large dots in the right plot) aligns most closely with the roofline estimates. The roofline models for the three GPUs are defined in \ref{app_tab:roofline_params}, while profiling numbers were extracted from AIConfigurator source code\cite{aiconfigurator}.
\begin{table}[ht]
    \centering

    \label{app_tab:roofline_params}
    \begin{tabular}{l|c|c}
        \toprule
        GPU & Peak compute (TOPS) & HBM bandwidth (TB/s) \\
        \midrule
        A100 & 624  & 2.0  \\
        H100 & 2000 & 3.35 \\
        B200 & 4500 & 8.0  \\
        \bottomrule
    \end{tabular}
        \caption{Peak compute throughput and memory bandwidth used in the roofline cost model.}
\end{table}
\paragraph{Isolated kernel profiling vs. end-to-end runs}
Finally, we verify that profiling kernels individually does not distort their latency compared to execution within the full model, where effects such as caching or fusion can lead to differences in the timings. Figure~\ref{app:fusion_gpu} shows that latencies obtained from isolated profiling closely match those extracted from end-to-end traces.
\section{Partition search algorithm}\label{app:algorithm}
Algorithm~\ref{alg:HELIX-dp} gives the pseudocode of the placement search based on dynamic programming described in Section \ref{ref:dp}.

\begin{figure*}[ht]
    \centering
     \includegraphics[width=0.925\textwidth]{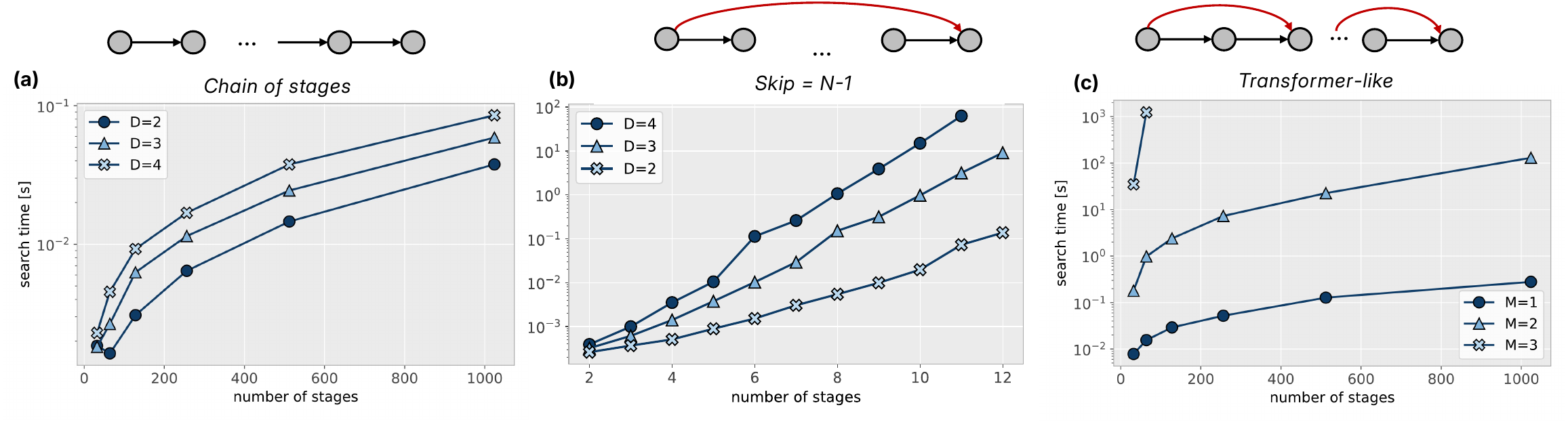}
\caption{HELIX-DP search time without memory tracking for (a) a chain of stages, (b) a chain with a residual skip from the first to the last stage (note the smaller range of $N$), and (c) a transformer-like DAG, with $D\in\{2,3,4\}$ heterogeneous devices.}
    \label{app_fig:algo_frontier_memory}
\end{figure*}

\paragraph{Preprocessing:} Memory is tracked only on the set $\mathcal{T}$ of devices whose capacity can be exceeded if every stage that lists them as a candidate were placed there. Stage footprints and capacities are expressed in units of the greatest common divisor (gcd) of the stages $b=\gcd_i m(s_i)$, rounding stage sizes up and capacities down, so each device's usage is a small integer.
\paragraph{Labels:} A label $\ell$ is a partial placement of $s_0,\dots,s_i$. It stores the memory usage $u^{\ell}$ on tracked devices, (ii) the placement of the \emph{frontier}, i.e.\ the stages whose outputs are still needed by unplaced consumers, together with the time each output within is available on each device, and (iii) the running latency $L^{\ell}_{\max}$ of the placed subgraph until the current stage. 
\paragraph{Label expansion and pruning:}
For each stage $s_i$, every label $\ell$ is expanded with every candidate device $d\in C_i$. 
If $d\in\mathcal{T}$ and placing $s_i$ on $d$ would exceed its capacity, the expanded label is discarded. 
Otherwise, the start time of $s_i$ is the latest arrival over its predecessors, each incurring a transfer cost when placed on a different device (Eq.~\ref{eq:dp-transition}), and its finish time updates $L_{\max}^{\ell'}$.
Stages with no remaining consumers are removed from the frontier.
Among labels with the same frontier placement, a label is discarded if another has no larger memory usage, running latency, and frontier arrival times.
If no label survives at some stage, the assignment is infeasible.
After the last stage, the label with minimal $L^{\ell}_{\max}$ gives the placement $a^{*}$.
\begin{algorithm}[t]
\caption{HELIX partitioning search}
\label{alg:HELIX-dp}
\begin{algorithmic}[1]
\REQUIRE DAG $G=(\mathcal{S},\mathcal{E})$, $C_i$, $T_{\mathrm{comp}}$, $T_{\mathrm{comm}}$, $m(s_i)$, $M_d$, $d_{\mathrm{init}}$
\ENSURE Placement $a^{*}$
\STATE $\mathcal{T}\gets\{d : \sum_{i:\,d\in C_i} m(s_i) > M_d\}$
\STATE $b\gets\gcd_i m(s_i)$
\STATE $\mathcal{L}\gets\{\ell_0\}$ with empty frontier, $u^{\ell_0}=\mathbf{0}$, $L^{\ell_0}_{\max}=0$
\FOR{$i=0,\dots,n-1$}
    \STATE $\mathcal{L}'\gets\emptyset$
    \FORALL{$\ell\in\mathcal{L}$ \textbf{and} $d\in C_i$}
        \IF{$d\in\mathcal{T}$ \textbf{and} $u^{\ell}_d+\lceil m(s_i)/b\rceil>\lfloor M_d/b\rfloor$}
            \STATE \textbf{continue}
        \ENDIF
        \STATE $\ell'\gets\ell\oplus(s_i\mapsto d)$
        \STATE $u^{\ell'}_d\gets u^{\ell}_d+\lceil m(s_i)/b\rceil$ \textbf{if} $d\in\mathcal{T}$
        \STATE $Q_{d'\to d}\gets A_{i-1}$ \textbf{if} $a_\ell(i-1)=d$ \textbf{else} $0$
        \STATE $L_i^{\ell'}(d)\gets$ Eq.~\ref{eq:dp-transition}
        \STATE $L_j^{\ell'}(d)\gets A_{i,a_\ell(j)}$ for each $j\in\mathrm{pred}(i)$ transferred to $d$
        \STATE $L^{\ell'}_{\max}\gets\max\big(L^{\ell}_{\max},\,L_i^{\ell'}(d)\big)$
        \STATE Remove from the frontier of $\ell'$ every $s_j$ with no consumer after $s_i$
        \STATE $\mathcal{L}'\gets\mathcal{L}'\cup\{\ell'\}$
    \ENDFOR
    \IF{$\mathcal{L}'=\emptyset$}
        \STATE \textbf{return} infeasible
    \ENDIF
    \STATE $\mathcal{L}\gets$ labels of $\mathcal{L}'$ not Pareto-dominated within their frontier assignment
\ENDFOR
\STATE $\ell^{*}\gets\arg\min_{\ell\in\mathcal{L}} L^{\ell}_{\max}$
\STATE \textbf{return} $a^{*}=a_{\ell^{*}}$
\end{algorithmic}
\end{algorithm}
\paragraph{Graph structure and number of devices.}
Figure~\ref{app_fig:algo_frontier_memory} reports the search time of HELIX-DP without tracking any memory devices ($|\mathcal{T}|=0$) on three synthetic DAGs: (a) a sequential chain, (b) a chain with a single residual skip from the first to the last stage, and (c) a transformer-like DAG with residual skips across every block. 
All devices can execute every stage, but they differ in compute throughput and memory bandwidth, so stage costs differ across devices and the search remains heterogeneous. 
We additionally vary the number of devices $D$ from 2 to 4.
For the chain and the transformer-like DAG, we sweep the number of stages $N$ from 32 to 1024. Search time grows roughly linearly with $N$, as expected from the $O(N|\mathcal{D}|^{\omega+1})$ bound, and stays below 0.4\,s at 1024 stages with four devices.
Although the transformer-like DAG contains many residual skips, each one closes at the next fan-in, within a couple of stages. Therefore $\omega$ is bounded by dependencies rather than by the depth of the model.
The long skip in (b) is the adversarial case. The frontier spans every stage between a residual producer and its fan-in, so a skip from the first to the last stage makes it cover the entire graph, and $\omega$ grows with $N$. The number of frontier keys then grows as $|\mathcal{D}|^{N}$, matching exhaustive search, so we restrict this sweep to $N\le12$. Search time increases by roughly a factor of $D$ per added stage, reaching $\sim$0.1\,s for $D=2$ and $\sim$9\,s for $D=3$ at 12 stages, and $\sim$60\,s for $D=4$ at only 11 stages. LLM DAGs avoid this regime, since their residual connections span only a few stages.
\begin{figure*}[ht!]
    \centering
    \includegraphics[width=0.957\textwidth]{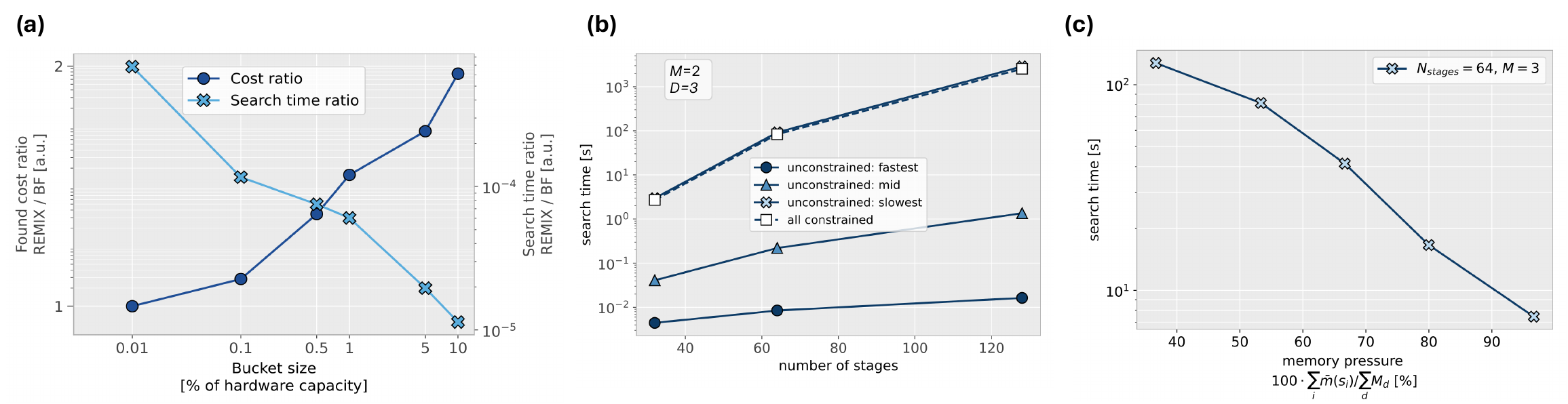}
    \caption{(a) Search time and accuracy of the found partition in terms of latency for HELIX DP versus exhaustive search (brute-force BF) depending on the bucket size, i.e. the unit on which the memory capacity is tracked for all devices. (b) Dependency of the search time for D=3 devices and $\mathcal{T}$ =2, varying on which of the devices remains out of the memory tracking (not overflowable or unconstrained) and (c) Dependency of the search time depending on how constrained the devices are with respect to the footprint of the DAG stages}
    \label{app_fig:mem_ablation}
\end{figure*}
\paragraph{Memory tracking.}
Figure~\ref{app_fig:mem_ablation}a evaluates the approximation introduced by tracking occupancy in buckets instead of bytes. We take the first five layers of Llama 8B with 5 stages each of different footprints (QKV, Attention, Out, Gate-Up, Down) and simulate the search for a batch of 32 queries at a context length of 16K tokens, and place them on three devices: an H100, 3DCIM and DRAM-PIM. Parameters and KV cache, both in 16-bit precision, create a DAG footprint of $\approx$ 12 GB of memory, and therefore both DRAM-PIM and H100 are set with a memory of 10 GB to force the tracking of their capacities. Exhaustive search (or brute force BF) runs over the same stages and conditions till termination, and therefore rovides the optimal placement.
We then run HELIX-DP with a bucket size $b$ ranging from 0.01\% to 10\% of device capacity. This means the memory is treacked in units of between 10 MB to 1 GB, whereas the smallest stage of this DAG (Out projection of 4096$\times$4096 parameters) has a $\approx$ 33 MB of footprint. 
We compare the search times and their reported latency of the found placement and report them both relative to BF. 
At $b=0.01\%$, HELIX-DP returns the same placement as BF with a search time over three orders of magnitude lower. 
Coarser buckets can reduce search time by roughly another order of magnitude, since more occupancy vectors coincide, which reduces the term $P$ from the complexity, meaning more labels are pruned by dominance.
This comes at the cost of rounding each footprint up to $\lceil m(s_i)/b\rceil$, which overestimates occupancy and the DP search ends up rejects feasible placements that could lead to the minimum latency. 
The latency of the found placement is to 1.08$\times$ slower than optimal solution retruned by exhaustive search at $b=0.1\%$, 1.46$\times$ at $b=1\%$, and close to 2$\times$ at $b=10\%$. This motivates setting $b$ to the greatest common divisor of the stage footprints (Section~\ref{ref:dp}), which represents every footprint exactly.
We follow the same testing settings as for Figure~\ref{app_fig:mem_ablation}a, but while varying the size of the DAG in terms of stages, memory capacities are adjusted proportionally to keep memory feasible placements. Figure~\ref{app_fig:mem_ablation}b fixes $D=3$ and $|\mathcal{T}|=2$ and varies which device is left untracked out of the three considered ones. Leaving the fastest device untracked keeps search time around $10^{-2}$\,s up to 128 stages, while leaving the slowest device outside the memory tracking is indistinguishable from tracking all three, reaching $\sim$2500\,s at 128 stages. 
The device with the lowest computational cost (i.e the fastest) for the stages dominates the placement, so its occupancy takes the largest number of distinct values and dominates the number of retained labels $P$. The slowest device receives few stages, so tracking it barely adds states. Search cost therefore highly depends on whether the devices that excel at the majority of stages in terms of execution cost, must be tracked.

Figure~\ref{app_fig:mem_ablation}c fixes $N=64$ and $|\mathcal{T}|=3$ and varies the memory pressure, defined as the aggregate stage footprint over the aggregate device capacity. This shows the dependency of the search time with how fast devices hit their memory capacity when searching through the graph in topological order. 
If a DAG has a footprint, for instance of 12 GB, and an admissible device $d$ has only 1 GB of capacity, the amount of labels that will involve $d$ will significanlty reduce compared to a case on which $M_d$ is 11.99 GB.
Search time drops from $\sim$130\,s at 37\% pressure to $\sim$8\,s at 96\%. 
Tighter capacities leave fewer reachable occupancy vectors feasible, so $P$ saturates earlier and more labels are discarded. 
The expensive regime is thus the loosely constrained one, where capacities are large relative to the footprint. 
In practice, this regime often makes tracking unnecessary altogether, since devices that cannot overflow are removed from $\mathcal{T}$.

\section{Computational primitives of evaluated models}
Table \ref{tab:nodes_vs_stages} shows the number of different computational primitives each of the evaluated models from the results section have, and therefore the ones that need pre-profiled costs in order to perform the partition search. 
Llama 8B and Mixtral 8x7B have 5 distinct primitives, i.e. placeable stages for the DAG construction, that repeat across all layers. 
Additionally, we also consider the first and final linear layers of the network (\texttt{embedding} and \texttt{lm\_head}). 
Bamba 9B instead alternates between mamba-like (\texttt{in}, \texttt{mixer} and \texttt{out}) and attention-like layers (\texttt{QKV}, \texttt{attn} and \texttt{out}), both followed by the standard FFN sequence of projections (\texttt{Gate-Up} and \texttt{Down}). 
\begin{table*}[h]
\centering
\label{tab:nodes_vs_stages}
\begin{tabular}{l|l|l|l|l}
\toprule
Model&\# of nodes&\# of Layers&\# of Primitives&No. of Stages\\
\midrule
\textbf{Llama-3.1 8B} & 2995 & 32 & 5 (\small QKV, Out, Attn, Gate-Up, Down) & 162 \\
\hline
\textbf{Bamba 9B} & 1784 & 32 & 8 (\small In, Mixer, Out | QKV, Attn, Out {\large $+$} Gate-Up, Down) & 162 \\
\hline
\textbf{Mixtral 8x7B} & 3043 & 32 & 5 (\small QKV, Out, Attn, Router, Experts) & 162 \\
\hline
GPT-OSS & 3455 & 36 & 5 (\small QKV, Out, Attn, Router, Experts) & 182 \\
\hline
Qwen3-Next-80B & 4525 & 48 & \begin{tabular}[t]{@{}l@{}}9 (\small QKVZ, BA, LinAttn, Out | QKV, Attn, Out\\ \small {\large $+$} Router, Experts)\end{tabular} & 278 \\
\hline
Kimi-Linear-48B & 2203 & 27 & \begin{tabular}[t]{@{}l@{}}12 (\small QKVZ, BA, LinAttn, F\_B, G\_B, Out |\\ \small KVA, QZ, MLA, Out {\large $+$} Router, Experts)\end{tabular} & 204 \\
\bottomrule
\end{tabular}
\end{table*}
Many newer models are still based on standard transformer-like patterns. 
They vary in their hidden dimensions, number of layers and number of experts, but can usually be summarized with the same computational primitives. 
A current trend is to replace attention blocks with more computationally efficient versions. 
For example, Qwen3-Next-80B and Kimi-Linear-48B extend the MoE structure and interleave linear attention layers with standard (GQA) or latent attention (MLA) layers, respectively. 
These newer attention blocks take a higher number of input tensors than the usual queries, keys and values, and therefore vary in the number of projections around the attention primitive. 
This represents further opportunities for HELIX to merge or split such parallel projections across hardware.

\section{3DCIM system simulator and backend}\label{3dcim_simulator}

The numbers for the 3DCIM platform are obtained via a simulator whose frontend is adapted to be registered as a custom backend for HELIX. With this frontend, the simulator can directly consume the \texttt graph captured by TorchDynamo when HELIX partitions the workload and allocates a partition to 3DCIM. To match individual operators to accelerator-level operators, the simulator transforms the given FX-level graph with a series of transformations based on pattern matching and replacement. For instance, linear layers involving individual slices of the Q, K, and V matrices are fused into a single linear layer with the combined QKV matrix, which can be efficiently computed on analog 3DCIM tiles. The transformed torch.fx graph is then lowered to a C++-based IR~\cite{Almaci2026GHL}, which manages lower-level operations such as tiling, batch splitting, data transfer modeling, and mapping. The resulting computational graph IR is scheduled, and the end time of the last operation is reported as the latency.

The 3DCIM platform system architecture is adopted from~\cite{Bchel2025Efficient}, which utilizes analog 3DCIM tiles of size \(512 \times 512\), connected to a landing pad with I/O buffers via a common bus. This bus conservatively blocks concurrent broadcasts and gathers across multiple activations. The bus transmits a tile-width's worth of inputs (512 bytes) in a bit serial manner with a startup latency of 50 cycles, according to internal circuit design findings. A total of 1024 3DCIM tiles are used to accommodate the weight capacity requirements of frontier LLM models. The MVM integration time on a 3DCIM tile is 200 ns. The greedy in-order weight mapping strategy from ~\cite{Bchel2025Efficient} is used, which maximizes tensor parallelism by distributing a single-stage MVM computation across as many tiles as possible. 3DCIM tiles are constrained by the one-tier-at-a-time constraint, which allows only one memory tier to be activated at a time. Since the weight mapping strategy places the experts of an FFN layer vertically across different tiers of 3DCIM tiles, our experiments for MoE-based workloads model serialized expert computation.

\section{Detailed HELIX partitions}
This section shows the full partitions found by HELIX for each heterogeneous system, with each stage coloured by the device it is mapped to. Their discussion can be found in the results section.
\begin{figure}[ht!]
    \centering
    \includegraphics[width=0.45\textwidth]{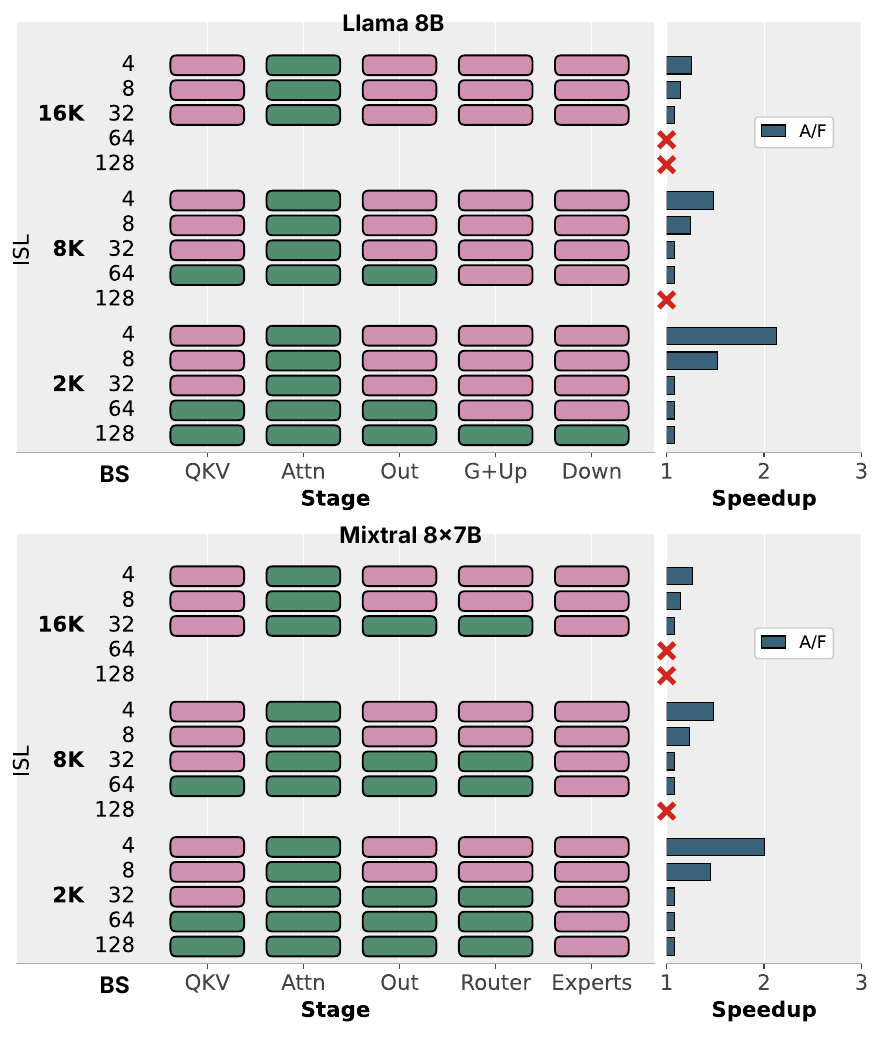}
    \vspace{-10pt}
    \caption{Partitions for 3DCIM + H100 for Llama 8B and Mixtral 8x7B}
    \label{app_fig:partitions_vs_afd2}
\end{figure}
\begin{figure}[ht!]
    \centering
    \includegraphics[width=0.45\textwidth]{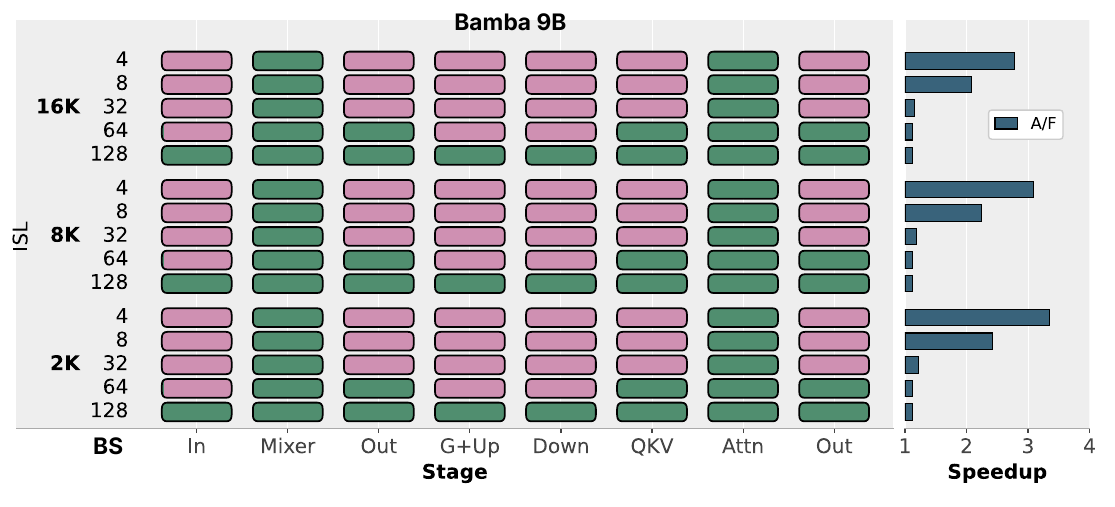}
    \vspace{-10pt}
    \caption{Partitions for 3DCIM + H100 for Bamba 9B}
    \label{app_fig:partitions_vs_afd}
\end{figure}

\paragraph{Partitions for H100 and + DRAM-PIM}: The AttAcc simulator~\cite{Park2024AttAcc} only models attention, so we estimate projection latencies on DRAM-PIM with a roofline model using AttAcc's bank-level, power-constrained configuration. Each HBM3 stack provides an internal bandwidth of $9\times670.4$\,GB/s, and the PIM units are provisioned at 1\,FLOP per byte, giving $\mathrm{BW}_{\mathrm{PIM}}\approx30$\,TB/s and $\mathrm{FLOPS}_{\mathrm{PIM}}\approx30$\,TFLOPS for 5 stacks. A projection with an $R\times C$ weight matrix applied to $V$ input vectors moves $\mathcal{B} = B\,(RC + VC + VR)$ bytes, with $B$ bytes per element, and performs $\mathcal{F} = 2VRC$ operations. Its latency is
\begin{equation*}
T_{\mathrm{PIM}} = \max\left(\frac{\mathcal{B}}{\mathrm{BW}_{\mathrm{PIM}}},\ \frac{\mathcal{F}}{\mathrm{FLOPS}_{\mathrm{PIM}}}\right).
\end{equation*}
In FP16, the arithmetic intensity of a projection is roughly $V$\,FLOP/byte, so it reaches the compute roof already at $V\approx1$ and its latency then grows linearly with $V$. In contrast, the H100 remains memory-bound up to a few hundred vectors. DRAM-PIM is therefore faster only while its $\sim$9$\times$ bandwidth advantage outweighs this linear growth, and the H100 outperforms it after a few concurrent queries (for our studies after batch size 8).
\begin{figure}[ht!]
    \centering    
    \includegraphics[width=0.45\textwidth]{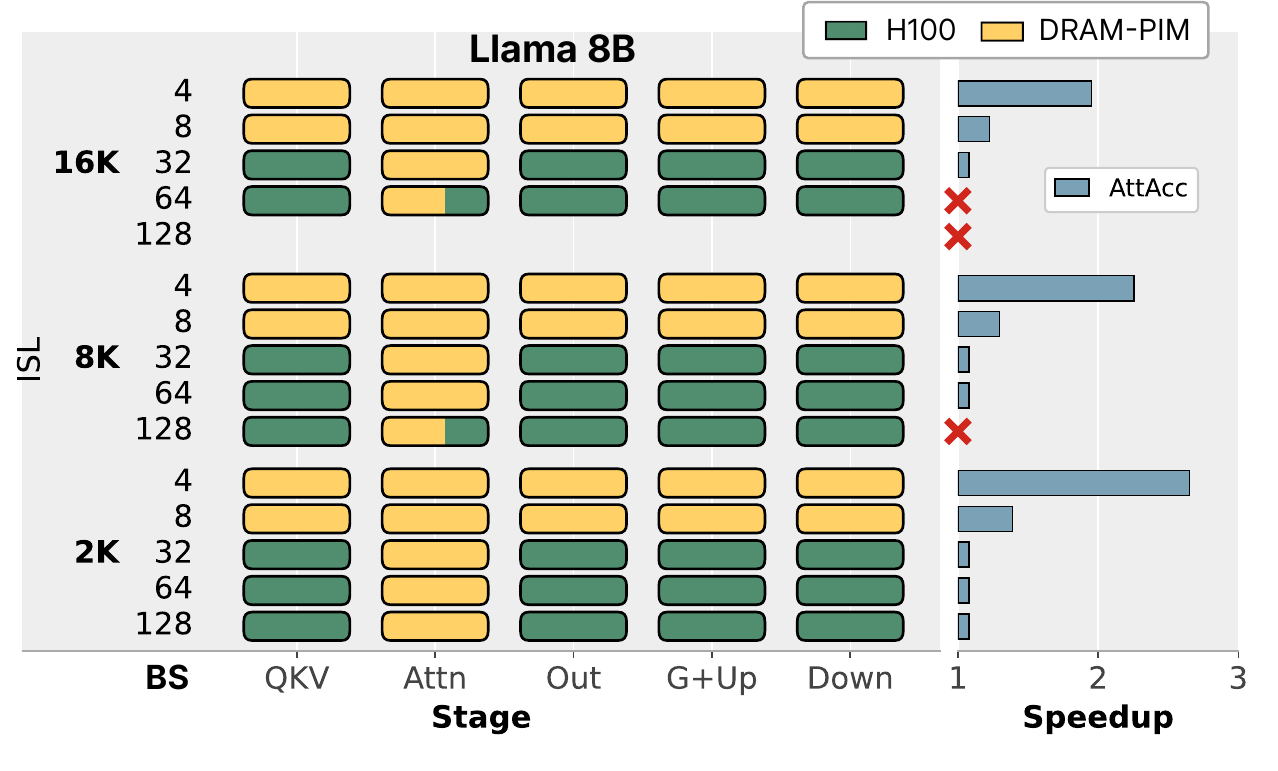}
    \vspace{-10pt}
    \caption{HELIX partitions for H100 + DRAM-PIM}
    \label{app_fig:partitions_vs_attacc1}
\end{figure}

\begin{figure}[h]
   \centering
\includegraphics[width=0.5\textwidth]{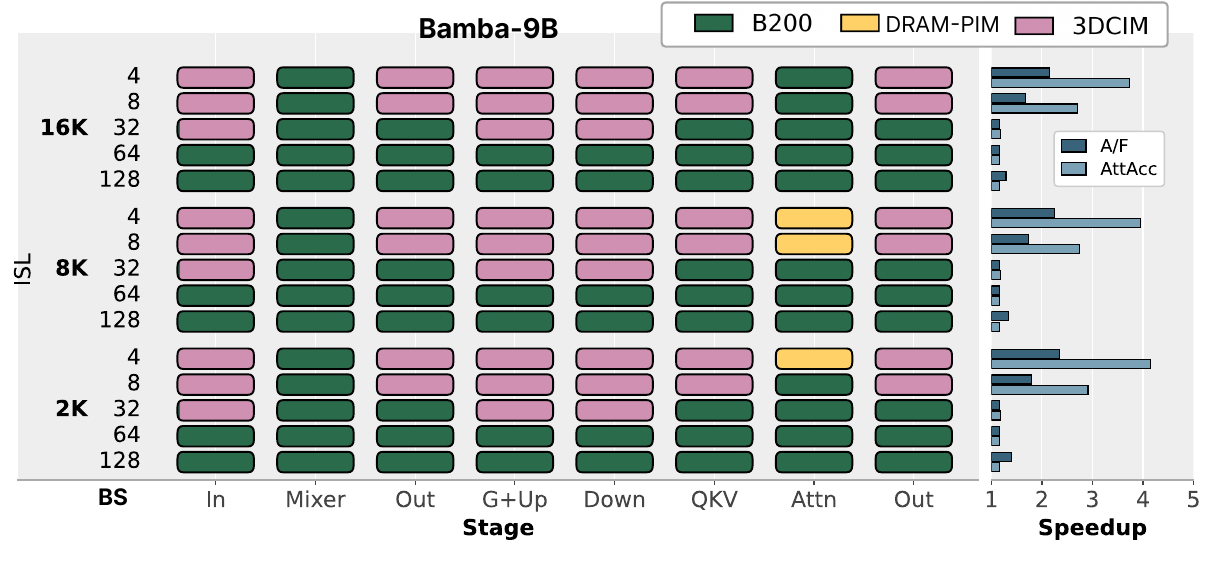}
\vspace{-10pt}
   \caption{HELIX partitions for B200 + 3DCIM + DRAM-PIM for Bamba}
    \label{fig:appx_3way_b2002}
\end{figure}

\begin{figure}[ht!]
   \centering
\includegraphics[width=0.45\textwidth]{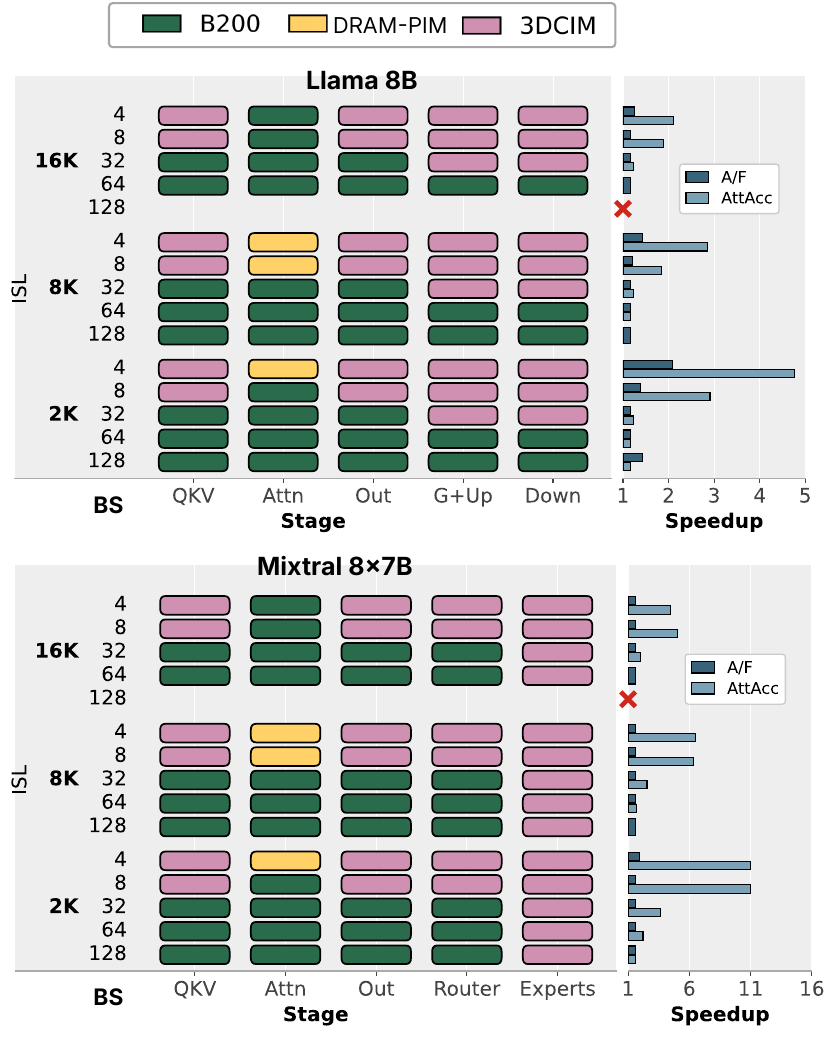}
\vspace{-10pt}
   \caption{HELIX partitions for B200 + 3DCIM + DRAM-PIM for Llama 8B and Mixtral 8x7B}
    \label{fig:appx_3way_b200}
\end{figure}
\end{document}